\documentclass[aps,amsmath,amssymb,twocolumn,showpacs]{revtex4-2}

\usepackage{graphicx}
\usepackage{amsmath}
\usepackage{bm}
\usepackage{amssymb}
\usepackage{bm}
\usepackage{bbold}
\usepackage{xcolor}
\usepackage{float}
\usepackage{chngcntr}
\usepackage{units}
\usepackage{dsfont}

\renewcommand{\vec}[1]{{\bm{#1}}}
\newcommand{\mat}[1]{{\underline{\bm{#1}}}}

\begin{document}

\title{Phononic heat conductance of gold atomic contacts:\\
Coherent versus incoherent transport}

\author{F. M\"{u}ller}
 \affiliation{Department of Physics, University of Konstanz, D-78457 Konstanz, Germany}
 
\author{J. C. Cuevas}
 \affiliation{Departamento de F\'{\i}sica Te\'orica de la Materia Condensada
and Condensed Matter Physics Center (IFIMAC), Universidad Aut\'onoma de Madrid, E-28049 Madrid, Spain}

\author{F. Pauly}
 \affiliation{Institute of Physics and Centre for Advanced Analytics and Predictive Sciences, University of Augsburg, D-86159 Augsburg, Germany}
  
\author{P. Nielaba}
 \affiliation{Department of Physics, University of Konstanz, D-78457 Konstanz, Germany}

\begin{abstract}

We present here a theoretical method to determine the phononic contribution to the thermal conductance 
of nanoscale systems in the phase-coherent regime. Our approach makes use of classical molecular dynamics (MD)
simulations to calculate the temperature-dependent dynamical matrix, and the phononic heat conductance is 
subsequently computed within the Landauer-B\"uttiker formalism with the help of nonequilibrium Green's function 
techniques. Tailored to nanostructures, crucial steps of force constant and heat transport calculations are performed 
directly in real space. As compared to conventional density functional theory (DFT) approaches, the advantage of our 
method is two-fold. First, interatomic interactions can be described with the method of choice. Semiempirical potentials 
may lead to large computational speedups, enabling the study of much larger systems. Second, the method naturally takes 
into account the temperature dependence of atomic force constants, an aspect that 
is ignored in typical static DFT-based calculations. We illustrate our method by analyzing the temperature 
dependence of the phononic thermal conductance of gold (Au) chains with lengths ranging from 1 to 12 atoms. Moreover, 
in order to evaluate the importance of anharmonic effects in these atomic-scale wires, we compare the phase-coherent 
approach with nonequilibrium MD (NEMD) simulations. We find that the predictions of the phase-coherent method and the 
classical NEMD approach largely agree above the Debye temperature for all studied chain lengths, which shows that heat 
transport is coherent and that our phase-coherent approach is well suited for such nanostructures.

\end{abstract}

\date{\today}

\maketitle

\section{Introduction} \label{introduction}

Understanding and controlling heat transport due to phonons in nanoscale systems and devices is of major 
importance in a variety of disciplines \cite{Pop2010,Cahill2014,Minnich2015}. Very recent experimental 
advances in nanothermometry have finally pushed thermal conductance measurements all the way down
to the scale of single atoms and molecules \cite{Cui2017,Mosso2017,Cui2019,Mosso2019}. Special attention has been paid 
to the case of metallic atomic-size contacts, which are known to be ideal systems to test basic theories of the 
transport of charge and energy at the nanoscale \cite{Agrait2003,Cuevas2017}. In this respect, 
numerous transport properties have been thoroughly analyzed in the context of these atomic-scale 
wires such as electrical conductance \cite{Krans1995,Scheer1998}, shot noise \cite{Brom1999,Wheeler2010,
Chen2012,Vardimon2016}, photocurrent \cite{Guhr2007,Viljas2007,Ittah2009,Ward2010}, thermopower
\cite{Ludoph1999,Tsutsui2013,Evangeli2015,Ofarim2016}, Joule heating \cite{Lee2013,Zotti2014}, or Peltier 
cooling \cite{Cui2018}, just to mention a few. 

The description of the heat conductance of metallic atomic contacts requires the use of fully atomistic 
methods \cite{Cui2017,Klockner2017,Burkle2018,Moehrle2019}. In particular, these methods have explained 
the observation of thermal conductance quantization in Au single-atom contacts and the fact that the 
Wiedemann-Franz law, which relates the electrical and thermal conductances in metallic systems, is 
approximately fulfilled irrespective of the size and material of the atomic contacts 
\cite{Cui2017,Klockner2017}. These theoretical approaches show for metallic atomic contacts that electrons largely 
determine the thermal conductance compared to phonons, similar to the situation in bulk metals. This insight was 
not obvious a priori because the transport mechanisms for fermionic electrons and bosonic phonons are 
fundamentally different in these nanoscale systems, as compared to bulk wires. Indeed, Ref.~\cite{Klockner2017} 
predicts that for aluminum single-atom contacts, phonon thermal transport may contribute as much as 40\% to the 
total thermal conductance, yielding a substantial deviation from the Wiedemann-Franz law. This interesting 
theoretical prediction, which presently lacks experimental confirmation, shows that these metallic nanowires are 
still very interesting from the point of view of phonon transport. 

Most theoretical calculations of the phonon contribution to the heat conductance 
of metallic atomic-size contacts assume that transport is fully coherent 
\cite{Cui2017,Klockner2017,Burkle2018}, i.e., that the inelastic mean free path for phonons is larger than 
the contact size. Other calculations \cite{Moehrle2019} use the classical Fourier's law, which includes anharmonic 
transport mechanisms. For single-atom contacts the coherent assumption appears to be reasonable. However, 
this is not obvious for longer and thicker junctions, since it is known that phonon mean free paths in metals can be as 
small as a few nanometers \cite{Jain2016}. Although the ab initio methods based on DFT, which are used so far 
\cite{Cui2017,Klockner2017,Burkle2018}, are very accurate, they are computationally demanding. Therefore their use is 
practically restricted  to relatively small contacts. In this respect, it would be useful to develop more efficient 
methods to describe the phonon transport in systems consisting of many atoms and to 
shed new light on the role of inelastic interactions. 

In this work we present an efficient method to compute the phononic heat conductance of any kind of nanojunction 
in the coherent transport regime. This method is based on the calculation of relevant atomic 
force constants, making use of classical MD simulations in thermal equilibrium, while the phonon transmission
and related heat conductance are computed in the framework of the Landauer-B\"{u}ttiker picture with the help of nonequilibrium Green's function
techniques. The approach detailed here considers the temperature dependence of the atomic force constants, 
which is in general important to describe the temperature dependence of the phononic heat conductance. We present 
a study of the phononic contribution to the thermal conductance of Au chains in the 
phase-coherent regime, systematically changing the length of the central chain of the junction. Moreover, we 
present a comparison of these calculations, which assume harmonic lattice distortions, with the results obtained using a recently 
developed procedure based on classical NEMD simulations that includes inelastic effects,
originating from anharmonic phonon-phonon scattering \cite{Moehrle2019}. This comparison allows us to test the accuracy 
of the phase-coherent approach for the description of phonon transport in these metallic nanoscale wires and provides 
insights into the regime, in which the harmonic description is valid. We will show that the coherent assumption 
is very accurate for all of the studied gold chains even for rather long ones, as the differences between 
the phase-coherent calculations and the NEMD are small at sufficiently high temperatures. This implies that phononic heat transport 
is coherent and that the ballistic assumption is valid for those atomic-size systems.

The rest of this paper is organized as follows. In Sec.~\ref{sec:basics}, we briefly describe the newly developed
computational method. We explain how we obtain the phononic thermal conductance with the help of nonequilibrium Green's function 
techniques and how the dynamical matrix is extracted from equilibrium MD. Furthermore, we 
recall in this section the basics of the classical NEMD approach, which we recently established \cite{Moehrle2019} to take into 
account anharmonic effects in phonon heat flow. We conclude Sec.~\ref{sec:basics} by discussing further technical details of our 
MD simulations and by presenting some test calculations. Section~\ref{sec:results} 
contains the main results of this work on the phononic heat conductance of Au single-atom contacts and 
chains, and in Sec.~\ref{sec:conclusions} we summarize our main findings. 

\section{Theoretical methods} \label{sec:basics}

In this section we describe in detail the two main transport methods that we employ later on to compute the 
phononic thermal conductance of atomic-scale metallic wires. We first discuss the novel coherent
transport method that is based on the determination of the dynamical matrix from classical MD simulations 
in thermal equilibrium. Then, we briefly review the NEMD procedure
that we have recently established to study phonon transport, taking into account anharmonic effects.
Finally, we conclude this section by presenting technical details of our MD simulations and some simple test calculations.

\subsection{Equilibrium molecular dynamics and phase-coherent phonon heat transport} \label{sec:dyn}

\begin{figure}[t]
\includegraphics[width=0.9\columnwidth,clip]{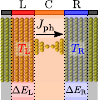}
\caption{(Color online) Typical geometry of a metallic atomic-size contact studied in this work. The contact is 
divided into three regions, namely the central scattering region (C, orange), the left reservoir (L, red) and the 
right reservoir (R, blue). In this example the contact consists of $1,784$ atoms in total, with 20 atoms located in the central part. 
On left and right sides two layers of fixed atoms (dark gray) stabilize the system during MD simulations. In the coherent 
transport description the temperatures $T_\text{L}$ and $T_\text{R}$ differ infinitesimally, and no inherent heat current 
occurs in the equilibrium MD runs. In NEMD with $T_\text{L}>T_\text{R}$ the phononic heat current $J_{\text{ph}}$ flows 
from the hot reservoir on the left at $T_\text{L}$ to the cold reservoir on the right at $T_\text{R}$. Energies 
added or subtracted to maintain a constant temperature gradient are denoted by $\Delta E_\text{L}$ and $\Delta E_\text{R}$.}
\label{fig:geo_NEMD}
\end{figure}

Let us start by describing, how the phononic heat conductance can be determined in the harmonic approximation and in the 
phase-coherent transport regime, where all the scattering events are elastic. Our goal is to compute the heat conductance due to the transport 
of phonons in a nanocontact, as the one depicted in Fig.~\ref{fig:geo_NEMD}. 

Within the Landauer-B\"uttiker approach for coherent transport the phononic heat conductance in the linear 
response regime is given by \cite{Mingo2003,Cuevas2017} 
\begin{equation}
\kappa_{\text{ph}} = \frac{1}{h}\int_0^\infty dE~E \tau_{\text{ph}}(E)\frac{\partial n(E,T)}{\partial T} ,
\label{eq:kappa}
\end{equation}
where $\tau_{\text{ph}}(E)$ is the energy-dependent phononic transmission function, $T$ is the temperature, and 
$n(E,T) = [\exp(E/k_{\rm B}T) -1]^{-1}$ is the Bose function. To compute the transmission $\tau_{\text{ph}}(E)$, we 
use standard nonequilibrium Green's function techniques \cite{Mingo2003,Cuevas2017}. The basic ingredient is the dynamical 
matrix $\mat{D}$, which we assume to be given. Its determination is discussed further below, and it is typically the most 
time-demanding computational step.

We start by dividing the atomic contact into three regions, as shown in Fig.~\ref{fig:geo_NEMD}: 
The left (L) and right (R) reservoirs, which are semiinfinite, and the central (C) part. Following this division, 
the dynamical matrix can be written as
\begin{equation}
	\mat{D} = 
	\begin{pmatrix}
	\mat{D}_\text{LL} & \mat{D}_\text{LC} & \mat{0}      \\
	\mat{D}_\text{CL} & \mat{D}_\text{CC} & \mat{D}_\text{CR} \\
	\mat{0}      & \mat{D}_\text{RC} & \mat{D}_\text{RR} 
	\end{pmatrix} ,
\label{eq:DynMat}
\end{equation}
where we assume that the reservoirs are not directly coupled. This latter assumption can typically be satisfied, by 
choosing the C region long enough. As discussed in Refs.~\cite{Mingo2003,Cuevas2017},
the phonon transmission function can be expressed in terms of the phonon Green's function $\mat{G}(E)$ as follows
\begin{equation}
\tau_{\text{ph}}(E) = \text{Tr}\left[\mat{G}_{\text{CC}}(E)\mat{\Lambda}_\text{L}(E)\mat{G}_{\text{CC}}^\dagger(E)
\mat{\Lambda}_\text{R}(E)\right] .
\label{eq:NEGF_tau}
\end{equation} 
Here, the retarded phonon Green's function of the C part is given by 
\begin{equation}
\mat{G}_{\text{CC}}(E) = \left[\left(E+\mathrm{i}\eta\right)^2\mathds{1}-\mat{D}_{\text{CC}}-\mat{\Sigma}_\text{L}(E)-
\mat{\Sigma}_\text{R}(E)\right]^{-1} ,
\label{eq:NEGF1}
\end{equation}
the embedding lead self-energies are
\begin{equation}
\mat{\Sigma}_Y(E) = \mat{D}_{\text{C}Y}\mat{g}_{YY}(E)\mat{D}_{Y\text{C}} ,
\label{eq:NEGF2}
\end{equation}
and $\mat{g}_{YY}(E)$ is the retarded surface Green's function of
the reservoir $Y= \text{L},\text{R}$. Finally, the broadening matrices $\mat{\Lambda}_Y(E)$, appearing in Eq.~(\ref{eq:NEGF_tau}), are related to the self-energies via
\begin{equation}
\mat{\Lambda}_Y(E) = \mathrm{i}\left[\mat{\Sigma}_Y(E) - \mat{\Sigma}^\dagger_Y(E)\right] .
\label{eq:NEGF3}
\end{equation}

The total transmission $\tau_{\text{ph}}(E)$ can be decomposed at each energy $E$ into contributions stemming from individual 
phonon transmission eigenchannels, in analogy to the electronic description \cite{Cuevas2017,Kloeckner2018}, as 
\begin{equation}
\tau_{\text{ph}}(E) = 
\mbox{Tr} \left[ \mat{t}_{\text{ph}}(E)\mat{t}^\dagger_{\text{ph}}(E) \right] = \sum_i \tau_{\text{ph},i}(E) ,
\end{equation}
where
\begin{equation}
\mat{t}_{\text{ph}}(E)=\mat{\Lambda}_\text{R}^{1/2}(E)\mat{G}_{\text{CC}}(E)\mat{\Lambda}_\text{L}^{1/2}(E) 
\end{equation}
is the transmission amplitude matrix and $\tau_{\text{ph},i}(E)$ are the corresponding eigenvalues of the
transmission probability matrix $\mat{t}_{\text{ph}}(E)\mat{t}^\dagger_{\text{ph}}(E)$.

With the knowledge of the dynamical matrix $\mat{D}$  we can thus compute the phononic heat conductance 
$\kappa_\text{ph}$ in the phase-coherent approximation. We now describe, how we obtain the dynamical matrix 
of a mechanical system by making use of classical MD in thermal equilibrium, assuming a harmonic 
model for the lattice vibrations \cite{Ashcroft1976}. 

We consider an infinite crystal in three spatial dimensions, where every atom $i$ is located at its 
equilibrium position $\vec{r}_i$. The corresponding displacement of the atom $i$ from the equilibrium position is specified by the 
vector $\vec{u}_i$. If the displacement is small for all $i$, we can use
the harmonic approximation and describe the potential energy of the crystal as follows
\begin{equation}
U = \frac{1}{2}\sum_{i\alpha,j\beta} \left( \frac{\partial^2U}{\partial u_{i\alpha} \partial 
u_{j\beta}} \right)_0 u_{i\alpha}u_{j\beta}.
\label{eq:U_taylor}
\end{equation}
Here $\alpha, \beta = \text{x}, \text{y}, \text{z}$ indicate the spatial direction, and the subscript $0$ reminds us that
the derivatives are evaluated at the equilibrium positions of the atoms $i, j$. For convenience
we introduce the following notation for the second derivatives of the potential 
\begin{equation}
\Phi_{i\alpha,j\beta} = \left( \frac{\partial^2 U}{\partial u_{i\alpha}\partial u_{j\beta}} \right)_0,
\label{eq:phi_allg}
\end{equation}
which are the atomic force constants. As usual, and due to the translational symmetry of a crystal, 
the diagonal elements ($i=j$) in the previous equation can be related to the off-diagonal elements ($i\neq j$) 
by the so-called acoustic sum rule
\begin{equation}
\Phi_{i\alpha,i\beta} = -\sum_{i\neq j}\Phi_{i\alpha,j\beta} .
\label{eq:phi_asr}
\end{equation}
The equation of motion of the atoms in this harmonic model reads
\begin{equation}
m_i\ddot{u}_{i\alpha} = -\sum_{j,\beta}\Phi_{i\alpha,j\beta}u_{j\beta} .
\label{eq:eq_of_motion}
\end{equation}
This relation has wave-like solutions of the form
\begin{equation}
u_{i\alpha}(\vec{r},t) = \frac{1}{\sqrt{m_i}}A_{i\alpha}
e^{i (\vec{q}\cdot\vec{r}-\omega t)} ,
\label{eq:waves}
\end{equation}
where $A_{i\alpha}$ is the amplitude and $\vec{q}$ the wave vector. Introducing this ansatz 
into Eq.~(\ref{eq:eq_of_motion}), we arrive at the secular equation 
\begin{equation}
\mat{D} - \omega^2\mathds{1} = 0 ,
\label{eq:dispersion}
\end{equation}
where
\begin{equation}
D_{i\alpha,j\beta} = \frac{1}{\sqrt{m_im_j}}\Phi_{i\alpha,j\beta}
\label{eq:Dyn}
\end{equation}
are the elements of the dynamical matrix and $m_i$ is the mass of atom $i$.

For the calculation of the dynamical matrix, we follow Refs.~\cite{Campana2006,Kong2009,Kong2011} 
and compute it by monitoring the displacement of the atoms in classical equilibrium MD simulations. 
In simple terms the central idea of this method can be understood as follows. Let us consider 
a single particle attached to a massless spring with spring constant $k$ and moving in one 
dimension along the x axis. The equipartition theorem relates the particle's elastic energy 
to the temperature $T$ (in the equipartition limit above the Debye temperature $T_\text{D}$) as follows
\begin{equation}
\frac{1}{2} k \langle u_\text{x}^2 \rangle = \frac{1}{2} m \langle v_\text{x}^2 \rangle =
\frac{1}{2}k_{\rm B} T ,
\end{equation}
which implies that 
\begin{equation}
k = \frac{k_{\rm B}T}{\langle u_\text{x}^2 \rangle} .
\label{eq:equi}
\end{equation}
Here, $u_\text{x}$ characterizes the fluctuations of the particle relative to its equilibrium position, 
$v_\text{x}$ is the velocity, and $\langle \dots \rangle$ denotes an ensemble average. This relation 
suggests that force constants can be determined as the inverse of mean square displacements.

To establish this connection we introduce the correlation matrix $\mat{K}$ that describes the pairwise 
correlations between atomic displacements, defined as \cite{Landau1980,Campana2006,Kong2011}
\begin{equation}
K_{i\alpha,j\beta} = \langle u_{i\alpha}u_{j\beta} \rangle =
\langle r_{i\alpha} r_{j\beta} \rangle - \langle r_{i\alpha} \rangle \langle r_{j\beta} \rangle .
\label{eq:displ}
\end{equation}
The dimension of the correlation matrix $\mat{K}$ is $3N\times3N$, if $N$ is the number of 
atoms in the system. Following the idea above and thanks to the equipartition theorem, we relate the force constants to the 
spatial correlations in Eq.~(\ref{eq:displ}) as follows 
\begin{equation}
\Phi_{i\alpha,j\beta} = k_{\rm B}T \left[\mat{K}^{-1}\right]_{i\alpha,j\beta} ,
\label{eq:phi}
\end{equation}
where $\left[\mat{K}^{-1}\right]_{i\alpha,j\beta}$ denotes the element $(i\alpha,j\beta)$ of the 
inverse of the correlation matrix $\mat{K}$. Finally, the dynamical matrix is obtained from 
Eq.~(\ref{eq:Dyn}). Since $\mat{D}$ is a thermodynamic quantity, it is affected by thermal 
noise in a finite sample. To take this into account, we introduce a cutoff $r_\text{c}$ such that if 
$\left|\vec{r}_i-\vec{r}_j\right| > r_\text{c}$, then $D_{i\alpha,j\beta}=0$ for our nanocontacts.

In detail, we calculate interatomic force constants of nanojunctions, see Fig.~\ref{fig:geo_NEMD}, directly in real 
space rather than in reciprocal space. This is adequate, because the atomic junctions do not possess any periodicity. The 
procedure delivers the dynamical matrix in Eq.~(\ref{eq:DynMat}), however with finite regions L and R. To compute the phonon 
transmission, we retain only the parts $\mat{D}_\text{CC}$, describing the central part, and the couplings to L and R electrodes, 
$\mat{D}_{\text{C}Y}=\mat{D}_{Y\text{C}}^\dagger$ with $Y=\text{L},\text{R}$. To improve the description of the electrodes, we 
perform separate calculations for bulk gold. For this purpose we employ periodic boundary conditions in three dimensions, construct 
the correlation matrix and subsequently the force constants in reciprocal space, as described in Refs.~\cite{Campana2006,Kong2009,Kong2011}. 
Through a Fourier transformation they are subsequently used to construct the semiinfinite electrodes in real space, effectively 
replacing the finite reservoir matrices $\mat{D}_\text{LL}$ and $\mat{D}_\text{RR}$ of the simulated nanojunctions with infinitely 
extended ones. It should however be noted that the electrodes enter only in terms of the electrode surface Green's functions in the 
calculation of the phononic transmission and derived thermal conductance. Therefore we compute electrode surface Green's functions 
with the help of the decimation technique from the bulk parameters \cite{Guinea1983}. The required length and minimal width of the 
surface region is determined by the couplings $\mat{D}_{\text{C}Y}=\mat{D}_{Y\text{C}}^\dagger$ of the nanojunctions. The procedure 
follows the idea of the cluster-based approach to electronic quantum transport, described in Ref.~\cite{Pauly2008}, where the nanojunction, 
simulated in real space, is considered as the extended central cluster.

Thus, we obtain the dynamical matrix from MD simulations at a certain temperature by monitoring the positions of the atoms in time. This 
holds as long as the system under study is in thermal equilibrium and the atoms fluctuate 
with small displacements around their equilibrium positions so that the harmonic approximation 
and equipartition theorem are valid. Let us stress that Refs.~\cite{Campana2006,Kong2009,Kong2011} calculate the correlation matrix and 
interatomic force constants in reciprocal space, as we do for the bulk parameters. However, the construction of the full correlation and 
force constant matrices directly in real space, as we perform it here for the atomic junctions of the form visible in Fig.~\ref{fig:geo_NEMD}, 
has not been implemented before to the best of our knowledge.

In the following we refer to the procedure of force constant determination through MD at a certain 
$T$ and subsequent heat conductance calculations in the Landauer-B\"uttiker framework, using Eq.~(\ref{eq:kappa}) 
at that same $T$, as MD@$T$-LB. An approximation along the lines of typical DFT approaches \cite{Buerkle2015} is 
to compute the force constants at a single fixed temperature $T_\text{fix}$, but to nonetheless determine $\kappa_{\text{ph}}$ at different $T$ 
through Eq.~(\ref{eq:kappa}). This assumes that force constants depend only weakly on temperature, and we refer to this simplified procedure as 
MD@$T_\text{fix}$-LB.

MD@$T$-LB and MD@$T_\text{fix}$-LB are both applicable to any system, where equilibrium MD simulations are available. 
The MD simulations can be based on empirical interatomic interactions to largely accelerate computations, but also ab-initio MD is possible.

\subsection{Nonequilibrium molecular dynamics and incoherent phonon heat transport}\label{sec:NEMD}

The method explained in the previous subsection describes phonon flow in the phase-coherent transport regime,
where anharmonic effects are assumed to be small. To assess the impact of inelastic phonon-phonon interactions on the 
phononic heat conductance of metallic atomic-size contacts beyond a temperature-dependent renormalization of harmonic 
force constants, we employ a method that we have recently presented \cite{Moehrle2019}. It is based on classical MD 
simulations under an applied temperature gradient, referred to as NEMD. We briefly describe it here, while details of the MD calculations 
are explained in the next subsection.

We consider an atomic contact, as depicted in Fig.~\ref{fig:geo_NEMD}, featuring a hot reservoir with temperature 
$T_\text{L}$ and a cold reservoir with temperature $T_\text{R} < T_\text{L}$. 
A phononic heat current $J_{\text{ph}}$ thus flows from left to right. Because we are only interested in the 
phonon transport through the central region, the atom dynamics of the reservoirs are of 
no direct importance. We therefore apply a strong rescale thermostat to both reservoirs. It rescales the velocities $v_i$ of the atoms, 
following the equipartition theorem
\begin{equation}
\langle E_Y \rangle = \sum_{i=1}^{N_Y} \frac{1}{2} m_i \langle v_i^2 \rangle = \frac{3}{2}N_Y k_{\rm B}T_Y ,
\label{eq:NEMD1}
\end{equation}
where $N_Y$ is the number of atoms in electrode $Y$. Rescaling to a target temperature $T_{Y,\text{t}}$ corresponds to a 
change $\Delta E_Y$ in energy of the thermostatted group of atoms
\begin{equation}
\Delta E_Y = \langle E_Y \rangle \left( 1-\frac{T_{Y,\text{t}}}{T_Y} \right).
\label{eq:NEMD2}
\end{equation}
Heating therefore results in a negative change in energy and cooling in a positive one, respectively. 
Following Fourier's law, the thermal conductance $\kappa_{\text{ph}}$ of the junction is given by
\begin{equation}
\kappa_{\text{ph}} = \frac{J_\text{ph}}{\Delta T} 
= \frac{1}{2\Delta t}
\frac{-\Delta E_\text{L} + \Delta E_\text{R}}{T_\text{L} - T_\text{R}} ,
\label{eq:NEMD3}
\end{equation}
Since the energy and temperature of the reservoirs fluctuate, we monitor $\Delta E_Y$ and $T_Y$ over a sufficient long 
time to reach reliable averages. 

Using classical statistics, the NEMD method takes into account both elastic and inelastic effects in phonon transport. Therefore 
a comparison of this method with the phase-coherent one, described above, will allow us 
to determine the relevance of inelastic scattering in the phonon transport through atomic contacts. We note that the NEMD 
description is expected to break down at sufficiently low temperatures, when quantum statistics play a role. Roughly, we expect 
that $T$ needs to be larger than the Debye temperature $T_\text{D}$ for the classical approximation to hold. The phase-coherent 
method in the harmonic approximation instead correctly incorporates Bose-Einstein statistics.

\subsection{Technical details of molecular dynamics simulations} \label{sec:sim}

The two transport methods described above use classical MD simulations, either in equilibrium or nonequilibrium. In this 
section we explain in detail how we carry out the simulations. 

All the MD simulations presented in this work are performed with the open-source package \textsc{Lammps} 
\cite{Plimpton1995,Lammps}. We describe interatomic interactions with the embedded-atom method 
\cite{Finnis1984}, taking the potential for Au from Ref.~\cite{Ackland1987}. The use of the same interatomic interaction potential 
for MD@$T$-LB, MD@$T_\text{fix}$-LB and NEMD ensures the comparability of the simulations. Junction geometries are oriented such 
that the $\langle 111 \rangle$ crystallographic direction coincides with the transport direction. The central scattering region of 
the gold contacts consists of two adjacent pyramids, connected by a chain of atoms. To the left and right of the central part, see 
Fig.~\ref{fig:geo_NEMD}, semiinfinite reservoirs with $630$ atoms are followed by two atomic layers of $252$ atoms with fixed atoms. 
These outermost regions are used to stabilize the junction during the simulations. For each junction geometry we did an energy minimization 
run for the C part, see Fig.~\ref{fig:geo_NEMD}, using the conjugate gradient method of \textsc{Lammps} prior to our MD simulations for 
force constant extraction. The reservoirs L and R are subsequently coupled to the equilibrated C part such that minimal pressure or stress 
is exerted on the studied junctions. Orthogonal to the transport direction we employ periodic boundary conditions to model extended surfaces 
of the reservoirs. The studied contacts with a variable number of chain atoms consist of around $1,800$ atoms in total. The length of the 
junctions varies from about $33$~{\AA} for the shortest contact up to $61$~{\AA} for the longest one, measured from the leftmost layer of reservoir 
L to the rightmost layer of reservoir R. The dimer contact geometry, shown in Fig.~\ref{fig:geo_NEMD}, exhibits a junction length of about 
$35$~{\AA}.

For statistical analysis we provide random initial velocities to the junction atoms in L, C, and R regions to obtain 
different time evolutions during the simulations. For each temperature and each geometry we perform 
20 individual simulations and average over the computed thermal conductance of this set. Thermal 
equilibration is achieved by applying a Nos\'{e}-Hoover thermostat to all junction atoms over a sufficiently long time of 1~ns to 
reduce fluctuations in temperature. This is done for the equilibrium MD and the NEMD simulations, respectively. 
If not mentioned otherwise, we use a time step of $1$~fs in the MD simulations. 

Following Sec.~\ref{sec:dyn}, the extraction of atomic force constants for the nanojunction in the coherent transport method is 
done in the NVE ensemble without an external thermostat to avoid interactions with an external heat bath. 
After the thermal equilibration of the system the ensemble average of Eq.~(\ref{eq:displ}) is taken over $200,000$ 
atomic configurations or frames to construct the correlation matrix, from which the force constants are obtained via Eq.~(\ref{eq:phi}). 
The total simulation time amounts to $20$~ns, with a snapshot being taken every $0.1$~ps to achieve a good sampling. 

Beside the force constants of the C part of the nanocontact and the couplings to L and R electrodes, we need them for the bulk-like 
electrodes. For this purpose, we simulate bulk Au using a primitive unit cell and periodic boundary conditions with a $\vec{k}$ grid 
of $16\times16\times 16$. Starting with randomized velocities, an equilibration is performed with a Nos\'{e}-Hoover thermostat for $1$~ns, 
as for the nanojunctions. Subsequently, correlations are determined in reciprocal space for an MD run in an NVE ensemble every $1$~ps during 
a total time interval of $20$~ns. Through the time average of the correlations, force constants are computed, which are then transformed to 
real space through a Fourier transformation, where we impose the FCC space group symmetry, as described for the electronic case in Ref.~\cite{Pauly2008}. 
Using the symmetrized bulk parameters in real space, electrode surface Green's function are constructed, assuming transverse periodic boundary 
conditions of $32\times 32$ lattice constants.

To account for the thermal noise in our Landauer-B\"uttiker-based transport calculations, we find it necessary to choose a finite cutoff radius 
$r_\text{c}$ for the atomic force constant determination of the nanocontacts, c.f.\ Sec.~\ref{sec:dyn}. To find out, which spatial cutoff is 
reasonable, we systematically studied the phononic thermal conductance as a function of $r_\text{c}$ for two junctions with chain lengths of $1$, 
the monomer, and $8$. The cutoff was varied from 2 to 52~{\AA} at $T=\unit[300]{K}$. For the monomer the phononic thermal conductance is basically 
independent of the cutoff for $\unit[6]{\AA}\le r_\text{c}\le\unit[14]{\AA}$, for the chain with a length of $8$ atoms $\kappa_\text{ph}$ is constant 
for $\unit[6]{\AA}\le r_\text{c}\le\unit[28]{\AA}$. Therefore, $r_\text{c}=10$~{\AA} appears to be a reasonable choice for short as well as long chains. 

The (targeted) temperature difference between the two reservoirs in the NEMD method is chosen as $\Delta T = T_\text{L} -T_\text{R} = 30$~K. The 
temperature difference is established by steadily heating the reservoir L, see Fig.~\ref{fig:geo_NEMD}, with a slope of $30$~K/ns up to the 
target temperature $T_\text{L,t}$, while the cold reservoir R is kept at $T_\text{R,t}$. The C part of the junction 
is simulated in the NVE ensemble without applying an external thermostat during this heating. After the heating process the two 
reservoirs L and R of the junction are once more equilibrated for $1$~ns at their respective temperatures $T_\text{L,t}$ 
and $T_\text{R,t}$ to reduce temperature fluctuations. 
Given the steadily maintained temperature difference, the changes of energy $\Delta E_Y$ and 
temperature $T_Y$ in the electrode $Y=\text{L},\text{R}$ are determined over a sufficient long time of $200$~ns, with a snapshot taken every $1$~ps 
to achieve a good sampling. For the calculation of the phononic thermal conductance value, we average over the last $50,000$ data points. Note that 
we always specify phononic conductance values of the NEMD calculations with respect to the target temperature $T_\text{R,t}$ of the cold reservoir.

Since we use a geometry, where the central part is directly 
coupled to two thermostatted electrodes, see Fig.~\ref{fig:geo_NEMD}, finite size effects need to be analyzed. Therefore, we systematically 
added $3$, $6$, $12$, $18$ and $36$ atom layers with the lateral size of those in L and R parts to the C part and studied 
the influence on the phononic thermal conductance. We found the thermal conductance to slightly increase with the size of the reservoirs, but the 
changes in $\kappa_\text{ph}$ were only on the order of a few pW/K at $\unit[300]{K}$. For this reason and for the sake of computational efficiency, 
we selected the contacts with the smallest number of atoms in the C region, as shown in Fig.~\ref{fig:geo_NEMD}. 

\subsection{Test of the phase-coherent transport method}\label{sec:tests}

\begin{figure}[t]
\includegraphics[width=\columnwidth]{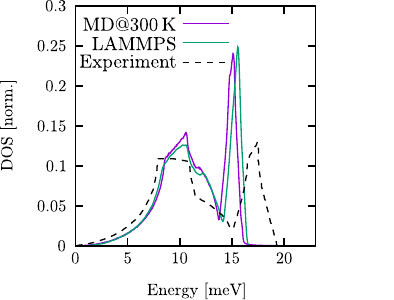}
\caption{(Color online) Bulk phonon DOS, calculated at 300~K (solid purple line), and 
experimental DOS (dashed black line) taken from Ref.~\cite{Lynn1973} at 297~K. Additionally we show a DOS determined at 
300~K with existing functionality of \textsc{Lammps}~\cite{Lammps,Kong2009,Kong2011} (solid green line). DOS curves are normalized by the total area enclosed.}
\label{fig:DOS}
\end{figure}

To test the accuracy of the phase-coherent transport method MD@$T$-LB, we first assess the quality of the bulk description. For this purpose we 
extracted bulk force constants in reciprocal space, as described in Secs.~\ref{sec:dyn} and \ref{sec:sim}, and compare the computed density of 
states (DOS) with available experimental data \cite{Lynn1973}. As it is visible in Fig.~\ref{fig:DOS}, the agreement is satisfactory. The main 
difference is that the theory underestimates the energy of the vibrational modes at the second peak of the DOS by about 20\%. The good agreement 
of our calculation with the DOS, computed with \textsc{Lammps} directly \cite{Lammps,Kong2009,Kong2011}, illustrates that deviations between theory 
and experiment originate from the employed potential itself. 

\begin{figure}[t]
\includegraphics[width=0.9\columnwidth]{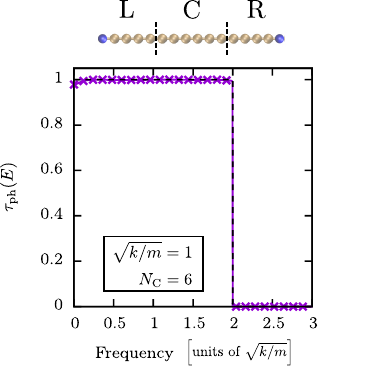}
\caption{(Color online) Phononic transmission function $\tau_{\text{ph}}(E)$ of a 1D chain with harmonic bonds. We compare the curve computed 
with force constants extracted from MD simulations (solid purple line and purple crosses) to the expected analytical behavior (dashed black line). 
The different quantities are given in reduced units. Above the plot we show the system under study, indicating L, C and R regions. The blue colored 
atoms are neighbors due to the assumed 1D periodic boundary conditions.}
\label{fig:plotTransChain}
\end{figure}

As another analytically trackable test case we consider a one-dimensional (1D) atomic chain. Employing MD within the NVE ensemble, we simulated 
a chain containing 16 atoms, assuming periodic boundary conditions, a mass of $m$ per atom, harmonic bonds with interatomic potentials 
$U=\frac{1}{2}k\left(r-r_0\right)^2$ and the spring constant $k$, see the inset of Fig.~\ref{fig:plotTransChain}. The equilibrium bond distance 
in the chain was set to $r_0$. Using a time step of $5 \times 10^{-3}\sqrt{m/k}$ in the MD simulations, the equilibration employed a Langevin 
thermostat at a temperature of $T=5 \times 10^{-3}kr_0^2/k_\text{B}$ for $5 \times10^6$ steps. Correlations were recorded regularly every $250$ 
steps for a total of $2 \times 10^7$ steps, from which we obtain the force constants and dynamical matrix. Note that we specify all physical 
quantities of this model in terms of reduced units of $r_0$, $m$, $k$ and $k_{\rm{B}}$. The 1D chain was divided into L, C and R parts, as shown 
in Fig.~\ref{fig:plotTransChain}, with the C part containing $N_\text{C}=6$ atoms. The phononic transmission $\tau_{\text{ph}}(E)$ was then determined 
within the nonequilibrium Green's function formalism, as described in Sec.~\ref{sec:dyn}, using the analytic solution for the surface Green's function 
of a semiinfinite chain with phononic nearest neighbor couplings $k$. As it is evident from Fig.~\ref{fig:plotTransChain}, our results compare in an 
excellent manner to the fully analytical solution that can be obtained for this simple model. The agreement holds in the whole range of energies 
studied, and especially the edge at $2\sqrt{k/m}$ is reproduced very accurately. 
\section{Results}\label{sec:results}

In this section we present the main results of this work. We analyze quasi-1D Au chains coupled to three-dimensional 
electrodes and discuss the influence of temperature and chain length on the phononic heat conductance. In addition, 
we compare phase-coherent transport to the NEMD method, which includes also inelastic effects from anharmonic phononic interactions.

\subsection{Length dependence of the phonon thermal conductance of Au chains}

The main strength of our phase-coherent transport method MD@$T$-LB is the inherent temperature 
dependence that MD simulations take into account and the time-efficient description even 
of large systems of several thousand atoms, if empirical interatomic interactions are used. We will demonstrate these 
aspects in this section by applying MD@$T$-LB and MD@$T_\text{fix}$-LB to quasi-1D Au chains. To analyze the importance 
of inelastic scattering events for the phononic thermal conductance, we consider different lengths of the central chain 
of the junction by varying atom numbers from $1$ to $12$. We will compare the predictions of the three different methods, 
introduced in Sec.~\ref{sec:basics}: (i) the phase-coherent method MD@$T$-LB, (ii) MD@$T_\text{fix}$-LB with $T_\text{fix}=100$~K, 
and (iii) classical NEMD. Method (ii) assumes that atomic force constants are basically independent of temperature and correspond 
to those at $100$~K. It is comparable to a static DFT ansatz, where force constants are for instance evaluated once with density 
functional perturbation theory for the ground state geometry \cite{Klockner2017,Kloeckner2018}.

\begin{figure*}[p!]
\centering
\includegraphics[width=0.9\textwidth]{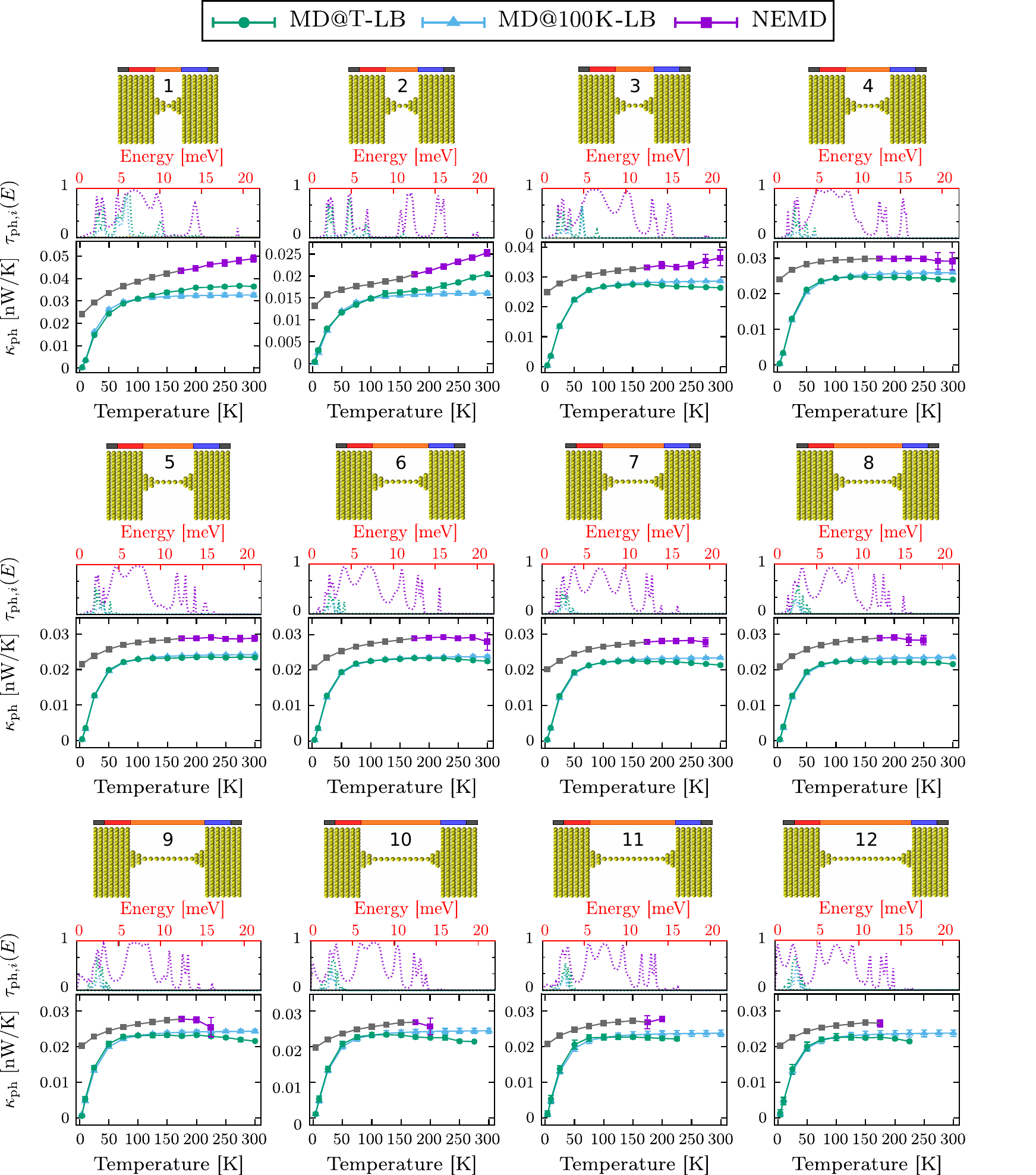}
\caption{(Color online) Phonon thermal transport through gold junctions. Junction geometries are shown in the top panels. 
The length of the central atomic chain ranges from 1 (top left) to 12 atoms (bottom right) between the two pyramids. Middle 
panels show phonon eigenchannel transmissions $\tau_{\text{ph},i}(E)$ with force constants determined at $100$~K. Phononic 
thermal conductance $\kappa_\text{ph}$, bottom panels, as a function of temperature for the three studied methods: We compare 
the results of the coherent method MD@$T$-LB (green circles), a simplified coherent method with a temperature-independent approximation 
for the dynamical matrix MD@100K-LB (blue triangles), and classical NEMD (purple squares). Mean values solely consider junctions that did 
not break during the 20 simulations performed at each $T$ in MD@$T$-LB and NEMD, and error bars specify standard deviations in this set of thermal 
conductances. For MD@100K-LB predictions of $\kappa_{\text{ph}}$ are made at each $T$ based on the force constants determined at 
$T_\text{fix}=100$~K, and error bars hence visualize the corresponding standard deviation. Since the NEMD is based on a classical 
theory, its results at sufficiently low temperatures are not meaningful. To visualize this, we grayed out data points for 
$T<T_\text{D}\approx170$~K.} 
\label{fig:results1}
\end{figure*}

Figure~\ref{fig:results1} shows the phononic thermal conductance as a function of temperature for the different 
computational methods. Going from the top left to the bottom right, the length of the central chain increases. The junction 
geometry is displayed in the top panels of Fig.~\ref{fig:results1}, middle panels show the transmissions of phonon eigenchannels 
as a function of energy for the phase-coherent calculation using force constants extracted at $100$~K, and bottom panels depict the phononic 
thermal conductance as a function of temperature. Values of $\kappa_\text{ph}$ represent averages over the 20 simulations performed in 
MD@$T$-LB and NEMD for each nanocontact at a fixed $T$, while error bars characterize the standard deviation. The averages solely consider 
junctions that did not break during the simulation runs. Since we find breakage especially for longer chains at elevated temperatures, the 
sample size for some data points is less than $20$. For example, for the junction with $12$ chain atoms, every NEMD run at or above a temperature 
of $200$~K broke before the end of the simulation time, and hence no data is available at $T\geq200$~K for this method. Similar statements hold 
for MD@$T$-LB, but we generally find that the metallic atomic contacts break at lower temperatures in NEMD than in equilibrium MD. Since we 
specify the temperature in NEMD with regard to the colder reservoir $T_\text{R,t}$, the applied temperature gradient leads to a higher effective 
temperature of the nanostructures in NEMD than in equilibrium MD. This rationalizes the reduced stability observed in the NEMD simulations. Since 
junctions are generally stable at $T=100$~K, MD@100K-LB yields data points throughout the entire temperature range studied. The standard deviations 
for MD@100K-LB characterize differences in predicted $\kappa_{\text{ph}}$ values, resulting from the force constants determined in the 20 simulation 
runs at 100~K. 

Concerning phonon transmission, Fig.~\ref{fig:results1} shows that for all junctions only up to $3$ transmission eigenchannels contribute. 
While the three channels may exhibit a similar transmission for energies $E \lesssim 7$~meV, a single channel determines the transmission 
above. 
For short chains, second and third channels range up to energies of $10$~meV but tend to vanish 
sooner for long chains. In general, a conduction channel is a linear combination of many different local modes. In the case of the studied chain junctions, 
the narrowest part of the wire is one-atom thick and can be assumed to be quasi 1D. In that limit, the conduction channels should reflect the phonon dispersion relation 
of a 1D chain, which is composed of three acoustic bands: Two transversal and one longitudinal phonon mode \cite{Klockner2017, Kloeckner2018}. This behavior is confirmed 
in our results (see middle panels of Fig.~\ref{fig:results1}) and conveys that we are mainly dealing with quasi 1D phonon transport.

Let us justify the choice of $T_\text{fix}=100$~K in MD@$T_\text{fix}$-LB for the approximation of the thermal conductance with 
temperature-independent force constants. MD@$T$-LB in Fig.~\ref{fig:results1} shows that the thermal conductance is well saturated 
at this temperature, particularly for long chains. Hence, $T_\text{fix}=100$~K allows us to identify crucial temperature-dependent changes 
in force constants by comparison of MD@$T$-LB and MD@100K-LB. Results of classical NEMD  are only valid at high enough temperatures 
$T\gtrsim T_\text{D}$. Above the Debye temperature $T_\text{D}$, the statistics of phonons are basically classical 
and quantum corrections can thus be neglected. For sufficiently low temperatures $T\ll T_\text{D}$, NEMD cannot predict the correct 
temperature dependence of the phononic thermal conductance. To signal caution, we have hence grayed out NEMD results for 
$T<T_\text{D}\approx 170$~K in Fig.~\ref{fig:results1}.

The thermal conductance predicted by MD@$T$-LB increases rather monotonically with $T$ in Fig.~\ref{fig:results1} for the short chains, 
especially the monomer and dimer, while we typically observe a saturation for longer chain lengths and subsequent weak decay. For the 
monomer and dimer, $\kappa_\text{ph}$ of MD@$T$-LB is larger for $T>100$~K compared to the values predicted with MD@100K-LB. This signals 
a significant dependence of force constants on temperature. Indeed, we assign the increase of the thermal conductance for short chain 
lengths and high temperatures to the thermal expansion of electrode reservoirs. Since we use fixed atoms to the left and right side, 
reservoirs can only expand towards the center of the junctions. This decreases the mean spatial distance between the atoms in 
the central scattering region and thus enhances the phononic thermal conductance. For longer chains, starting with the trimer, MD@$T$-LB 
and MD@100K-LB give very comparable results. Indeed the heat conductance predicted by MD@100K-LB overestimates those of MD@$T$-LB, which 
slightly decays at high enough $T\gtrsim 150$~K. We assign this effect to disordered chains, which form due to the thermal expansion of the 
electrode reservoirs. If we monitor the mean displacement 
of the chain atoms transverse to the transport direction, the chains are less 
linear at higher temperature, increasing the scattering and thus lowering the overall phononic transmission. In that sense, the temperature 
affects the properties of the phonon modes and subsequently the phononic transmission as the geometry of the central chains slightly changes.
The small deviations of a few pW/K between MD@$T$-LB and 
MD@100K-LB methods show that a temperature-independent description of atomic force constants is generally a good approximation for the 
studied systems. Differences between MD@$T$-LB and MD@100K-LB may be used to learn more about the mechanical and thermal characteristics of a nanostructure.

The heat conductance determined with NEMD in Fig.~\ref{fig:results1} is always larger than those of MD@$T$-LB. Quantitatively, within the 
limits of applicability of classical statistics, i.e.\ $T\gtrsim T_\text{D}$, deviations are relatively small and at most of the order of 
10~pW/K. The qualitative behavior agrees very well, i.e., $\kappa_\text{ph}$ increases rather monotonically with $T$ for short chains, 
including the monomer, dimer and additionally the trimer,  while we observe a saturation for longer chain lengths and a slight decay. 
The suppression of the heat conductance in NEMD for long chain length and high $T$ could in principle be assigned both to the effect of 
increasing chain disorder by thermal reservoir expansion and increasing inelastic scattering due to phonon-phonon interactions 
\cite{Mingo2006}. However, the quantitatively similar trends of the phase-coherent method MD@$T$-LB suggest that the slight suppression 
of $\kappa_\text{ph}$ at high $T$ originates from elastic scattering in an effectively disordered chain. 

Overall, the agreement of results of the classical NEMD and the temperature-dependent, coherent method MD@$T$-LB in Fig.~\ref{fig:results1} 
is very good. We note that we are comparing two methods that are making completely different assumptions about the theoretical 
description of phononic heat transport. Let us also emphasize that our computational results are statistically very stable, as can be seen by 
the small error bars. Absolute differences are only about a few pW/K, and we observe that differences between NEMD and MD@$T$-LB tend to 
decrease with increasing chain length. While we use a high sample size in our calculations and large nanocontacts, differences may still arise 
from systematic numerical uncertainties of our finite-size study.

The phononic thermal conductance of single-atom contacts has already been studied and discussed 
intensively in other works using DFT-based methods \cite{Cui2017, Klockner2017, Burkle2018}. In 
Ref.~\cite{Cui2017} phononic thermal transport was found to yield only about 4\% of the 
total thermal conductance $\kappa=\kappa_\text{el}+\kappa_\text{ph}$ at room-temperature for a gold dimer 
junction. With a prevalent electronic contribution of $\kappa_\text{el}\approx 0.59$~nW/K this corresponds to a phononic 
thermal conductance of about $\kappa_\text{ph}\approx\unit[0.025]{nW/K}$ at $T=300$~K. Our results 
predict a phononic thermal conductance of $\kappa_\text{ph}=0.020$~nW/K at $T=300$~K for the studied 
gold dimer junction in Fig.~\ref{fig:results1} using the MD@$T$-LB approach, which is in excellent agreement 
with the one calculated via DFT \cite{Cui2017}. Another DFT-based calculation of a gold dimer was presented in 
Ref.~\cite{Klockner2017}, where the validity of the Wiedemann-Franz law was investigated for metallic atomic-size 
contacts. In that case Kl\"{o}ckner \textit{et al.} determined a phononic thermal conductance of 
$\kappa_\text{ph}=0.051$~nW/K at room-temperature, which is about $2.5$ times higher than our result in 
this work. In both reports~\cite{Cui2017, Klockner2017}, the geometry of the studied dimer junction slightly differs 
from the one that we are using here. However, these comparisons show that our MD@$T$-LB approach is a valid description of 
phononic thermal transport. 

\begin{figure}[t]
\includegraphics[width=1.0\columnwidth]{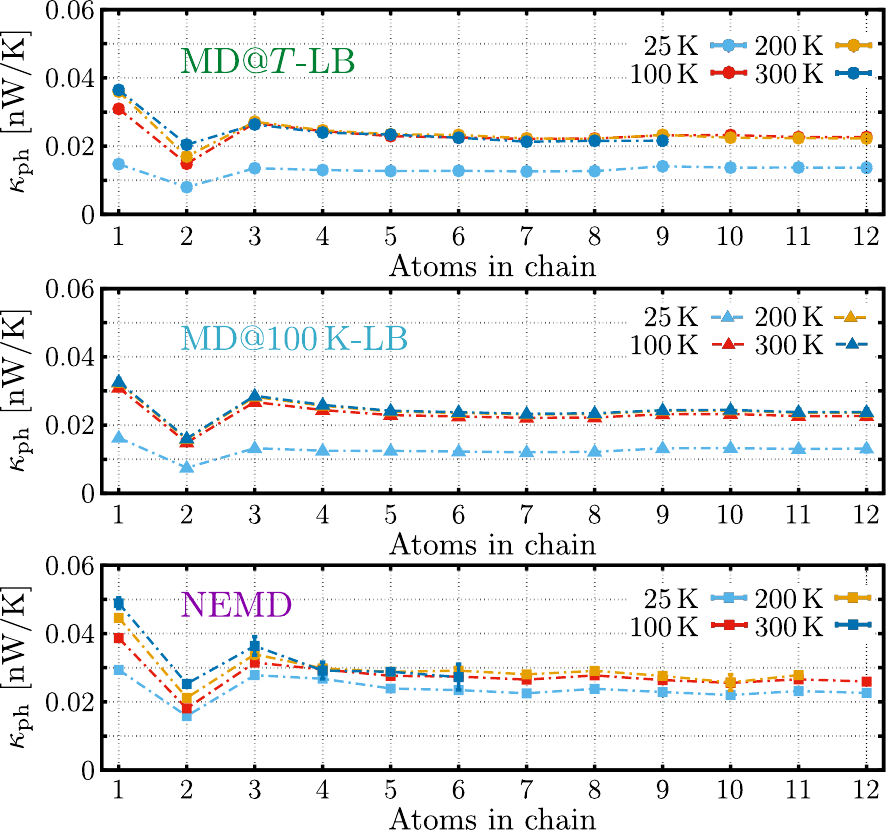}
	\caption{(Color online) Thermal conductance $\kappa_\text{ph}$ as a function of the number of atoms 
	in the chain at temperatures of $25$, $100$, $200$ and $300$~K. The results of the three 
	different methods are shown separately, namely for MD@$T$-LB, MD@100K-LB and 
	NEMD, going from top to bottom.}
\label{fig:results2}
\end{figure}

To get a more detailed view on the length dependence of the thermal conductance, we show it as a function of the number of atoms 
in the central chain of the junction for selected temperatures in Fig.~\ref{fig:results2}. Results of the three studied 
methods are compiled in separate panels. In every case, we see finite-size oscillations and a decline of $\kappa_\text{ph}$ 
with increasing chain length. For larger chain lengths, $\kappa_\text{ph}$ is rather constant and shows only weak 
oscillations. For all three methods, the thermal conductance is suppressed at a chain length of two atoms. Although we are not certain about the origin of 
this suppression, this reduction of the thermal conductance can be traced back to a suppression of the phononic transmission of the dominant conduction channel in the 
energy range between $5$ and $10$~meV (see the corresponding panel containing the phononic transmission coefficients for the dimer junction in Fig.~\ref{fig:results1}). 
This suggests a destructive interference, which may be related to a specific symmetry of the phonon modes giving rise to this conduction channel 
(mainly transversal modes as shown in Ref.~\cite{Kloeckner2018}) and the coupling of those phonons to the leads. However, whether this effect is real or an 
artifact of the employed embedded-atom method potential is unclear and remains as an interesting point for experimental examination. The plots 
for MD@$T$-LB and MD@100K-LB also show that above about $100$~K the thermal conductance depends only weakly on temperature. For the thermal conductance 
of the classical NEMD calculations in the bottom panel of Fig.~\ref{fig:results2} we notice only a weak temperature dependence. Consistent with 
Fig.~\ref{fig:results1}, the thermal conductance at $T=25$~K is very similar to those at $T=100$~K. This erroneous prediction arises from 
classical statistics and differs from MD@$T$-LB and MD@100K-LB, which both properly take Bose-Einstein statistics into account. We note that 
some data points are missing from MD@$T$-LB and NEMD, since junctions with a long atomic chains tend to break at elevated temperatures, as 
already discussed in the context of Fig.~\ref{fig:results1}. 

DFT-based calculations of atomic junctions of Au for chain lengths ranging from two to five atoms by B\"{u}rkle and Asai~\cite{Burkle2018} 
show that the ratio $\kappa_\text{ph}/\kappa_\text{el}$ between the phononic and the electronic thermal conductance at room-temperature 
depends only weakly on the length of the junction. Apart perhaps from the data point of the dimer, this is consistent with the results 
of Fig.~\ref{fig:results2} in the sense that the phononic thermal conductance is almost independent of the chain length.

To summarize, for $T\gtrsim T_\text{D}$ the results of the three different calculation methods differ only by a few pW/K for Au metallic atomic contacts containing 
long chains. This strongly suggests that the phononic thermal transport through the studied junctions is ballistic even for rather long chains 
with up to $12$ atoms. Longer chains could not be studied here for practical reasons, since most of them broke during simulations. While it 
would be interesting to investigate much longer chains to see if there is a critical length above which transport is no longer elastic, 
experiments indicate that a maximum length of up to eight chain atoms can be realized \cite{Smit2001,Yanson1998,Agrait2002}. Altogether our 
results corroborate that the harmonic approximation used in the coherent transport calculations is well suited for the studied atomic-size 
junctions. 

\subsection{Computational costs}

To evaluate the efficiency of the newly introduced phase-coherent phonon transport methods MD@$T$-LB and MD@$T_\text{fix}$-LB, let us now compare 
the computational costs of the presented simulation procedures. We do this exemplarily for the monomer junction shown in the upper leftmost panel 
of Fig.~\ref{fig:results1}. The junction consists of $1,783$ atoms. 

At a fixed temperature we need about $288$ core-h for one NEMD run, including the MD simulation and calculation of the phononic thermal conductance. 
With a total number of $20$ individual runs per temperature to construct a sufficient statistical data base and $14$ 
different temperatures, this value needs to be multiplied by a factor of $280$. The study of a single junction using NEMD, as 
shown for $\kappa_\text{ph}$ in the upper leftmost panel of Fig.~\ref{fig:results1}, thus corresponds to a computational cost of about $80,640$ core-h. 

In the phase-coherent regime, using MD@$T$-LB, the calculation of the phononic thermal conductance takes about $48$ core-h. This includes $28$ core-h for 
the equilibrium MD simulation and additional $20$ core-h for the calculation of force constants and transport. 
With the same statistical sample of $20$ individual simulations per temperature and $14$ different studied temperatures, this results 
in a total computational cost of about $13,440$ core-h. This is about $6$ times faster than NEMD. If we only compare the MD part of the calculations, 
the coherent description is about $10$ times faster than NEMD. 

If we disregard the temperature dependence of the interatomic force constants and use the simplified MD@$T_\text{fix}$-LB for the evaluation of 
phase-coherent phononic conduction, a single geometry can be 
studied even more efficiently. As for MD@$T$-LB, the total computational cost of a single transport calculation at $T_\text{fix}=100$~K equals to 
about $48$ core-h. With $20$ individual simulations at this temperature for statistical relevance, this totals to about $960$ core-h for a single 
junction geometry. Note that the transport calculations at the remaining 13 different temperatures correspond to negligible computational demand, 
which we neglect here. Hence, MD@$T_\text{fix}$-LB requires about a factor 14 less core-h as compared to MD@$T$-LB and a factor of 84 as compared 
to NEMD. Considering that results of MD@$T$-LB and MD@$T_\text{fix}$-LB differ little for the metallic atomic contacts and in relation to the 
computational costs, this approximation can certainly be used to obtain an excellent first guess. 

\section{Conclusions} \label{sec:conclusions}

In this work we have introduced a method to calculate atomic force constants of nanostructures directly from MD simulations in thermal equilibrium 
by tracking atomic positions in real space. The force constants are then used to evaluate the phononic contribution to the thermal conductance of 
nanocontacts. This step utilizes the framework of the Landauer-B\"{u}ttiker scattering theory for coherent transport, expressed in terms of 
nonequilibrium Green's function techniques. At the example of monoatomic metallic nanocontacts we have confirmed the validity and accuracy of 
this novel approach. 

The advantage of the new coherent transport method is two-fold. On the one hand, if reliable semiempirical interatomic interaction potentials 
are available, it is less time-consuming than ab initio approaches, which facilitates the modeling of large systems. We have illustrated this fact 
by showing calculations for junctions with up to 1794 atoms, which are difficult to handle with conventional DFT-based methods. 
On the other hand, our MD-based approach automatically takes into account the temperature dependence of the dynamical matrix, which is usually 
ignored in static ab initio procedures.  

For the gold metallic nanowires, featuring lengths of atomic chains from $1$ to $12$ atoms, we have assessed the relevance of inelastic effects 
due to anharmonic phonon-phonon scattering by comparing results of the phase-coherent method to those obtained with classical NEMD. While being 
applicable only above the Debye temperature because of classical statistics, the latter approach naturally takes inelastic effects into account. 
At sufficiently high temperatures (i.e. above the Debye temperature), where all relevant phonon modes are fully occupied, the phononic thermal 
conductance is only weakly dependent on the temperature and phononic transport is no longer very sensitive to an increase 
in temperature. In that sense, the saturation of the phononic thermal conductance at elevated temperatures, which we observed for the MD@$T$-LB as well as 
the NEMD description, is a clear signature of the ballistic nature of the phonon conduction in the studied metallic atomic-size junctions.
Overall, our analysis of the temperature and length dependence of the phononic heat conductance shows that transport through these nanostructures 
is indeed coherent. 

To conclude, we emphasize that the coherent transport method, introduced in this work, is by no means restricted to metallic atomic-size contacts: 
It can be applied to describe the coherent phonon transport in any nanostructure, if classical MD simulations can be carried out. In this sense we 
anticipate that this method may become very useful to investigate different questions related to heat flow through molecular junctions 
\cite{Cui2019,Mosso2019}, where the thermal conductance is determined by phonon conduction.

\section{Acknowledgment} \label{Acknowledgement}

We thank J.C.\ Kl\"ockner for many stimulating discussions. P.N.\ and F.P.\ were funded by the Collaborative 
Research Center (SFB) 767 of the German Research Foundation (DFG). J.C.C.\ thanks the Spanish Ministry of 
Science and Innovation (Grant No. PID2020-114880GB-I00) for financial support as well as the DFG and SFB 
767 for sponsoring his stay at the University of Konstanz as a Mercator Fellow. F.M.\ and P.N.\ gratefully 
acknowledge the Gauss Centre for Supercomputing e.V.\ (www.gauss-centre.eu) for providing computing time 
through the John von Neumann Institute for Computing (NIC) on the Supercomputer JUWELS at J\"ulich 
Supercomputing Centre (JSC). Furthermore, the authors acknowledge support by the state of Baden-W\"urttemberg 
through the bwHPC initiative.

%

\begin{thebibliography}{49}%
\makeatletter
\providecommand \@ifxundefined [1]{%
 \@ifx{#1\undefined}
}%
\providecommand \@ifnum [1]{%
 \ifnum #1\expandafter \@firstoftwo
 \else \expandafter \@secondoftwo
 \fi
}%
\providecommand \@ifx [1]{%
 \ifx #1\expandafter \@firstoftwo
 \else \expandafter \@secondoftwo
 \fi
}%
\providecommand \natexlab [1]{#1}%
\providecommand \emph  [1]{``#1''}%
\providecommand \bibnamefont  [1]{#1}%
\providecommand \bibfnamefont [1]{#1}%
\providecommand \citenamefont [1]{#1}%
\providecommand \href@noop [0]{\@secondoftwo}%
\providecommand \href [0]{\begingroup \@sanitize@url \@href}%
\providecommand \@href[1]{\@@startlink{#1}\@@href}%
\providecommand \@@href[1]{\endgroup#1\@@endlink}%
\providecommand \@sanitize@url [0]{\catcode `\\12\catcode `\$12\catcode
  `\&12\catcode `\#12\catcode `\^12\catcode `\_12\catcode `\%12\relax}%
\providecommand \@@startlink[1]{}%
\providecommand \@@endlink[0]{}%
\providecommand \url  [0]{\begingroup\@sanitize@url \@url }%
\providecommand \@url [1]{\endgroup\@href {#1}{\urlprefix }}%
\providecommand \urlprefix  [0]{URL }%
\providecommand \Eprint [0]{\href }%
\providecommand \doibase [0]{http://dx.doi.org/}%
\providecommand \selectlanguage [0]{\@gobble}%
\providecommand \bibinfo  [0]{\@secondoftwo}%
\providecommand \bibfield  [0]{\@secondoftwo}%
\providecommand \translation [1]{[#1]}%
\providecommand \BibitemOpen [0]{}%
\providecommand \bibitemStop [0]{}%
\providecommand \bibitemNoStop [0]{.\EOS\space}%
\providecommand \EOS [0]{\spacefactor3000\relax}%
\providecommand \BibitemShut  [1]{\csname bibitem#1\endcsname}%
\let\auto@bib@innerbib\@empty
\bibitem [{\citenamefont {Pop}(2010)}]{Pop2010}%
  \BibitemOpen
  \bibfield  {author} {\bibinfo {author} {\bibfnamefont {E.}~\bibnamefont
  {Pop}},\ }\bibfield  {title} {\emph {\bibinfo {title} {Energy dissipation and
  transport in nanoscale devices},}\ }\href@noop {} {\bibfield  {journal}
  {\bibinfo  {journal} {Nano Res.}\ }\textbf {\bibinfo {volume} {3}},\ \bibinfo
  {pages} {147} (\bibinfo {year} {2010})}\BibitemShut {NoStop}%
\bibitem [{\citenamefont {Cahill.}\ \emph {et~al.}(2014)\citenamefont
  {Cahill.}, \citenamefont {Braun}, \citenamefont {Chen}, \citenamefont
  {Clarke}, \citenamefont {Fan}, \citenamefont {Goodson}, \citenamefont
  {Keblinski}, \citenamefont {King}, \citenamefont {Mahan}, \citenamefont
  {Majumdar}, \citenamefont {Maris}, \citenamefont {Phillpot}, \citenamefont
  {Pop},\ and\ \citenamefont {Shi}}]{Cahill2014}%
  \BibitemOpen
  \bibfield  {author} {\bibinfo {author} {\bibfnamefont {D.~G.}\ \bibnamefont
  {Cahill.}}, \bibinfo {author} {\bibfnamefont {P.~V.}\ \bibnamefont {Braun}},
  \bibinfo {author} {\bibfnamefont {G.}~\bibnamefont {Chen}}, \bibinfo {author}
  {\bibfnamefont {D.~R.}\ \bibnamefont {Clarke}}, \bibinfo {author}
  {\bibfnamefont {S.}~\bibnamefont {Fan}}, \bibinfo {author} {\bibfnamefont
  {K.~E.}\ \bibnamefont {Goodson}}, \bibinfo {author} {\bibfnamefont
  {P.}~\bibnamefont {Keblinski}}, \bibinfo {author} {\bibfnamefont {W.~P.}\
  \bibnamefont {King}}, \bibinfo {author} {\bibfnamefont {G.~D.}\ \bibnamefont
  {Mahan}}, \bibinfo {author} {\bibfnamefont {A.}~\bibnamefont {Majumdar}},
  \bibinfo {author} {\bibfnamefont {H.~J.}\ \bibnamefont {Maris}}, \bibinfo
  {author} {\bibfnamefont {S.~R.}\ \bibnamefont {Phillpot}}, \bibinfo {author}
  {\bibfnamefont {E.}~\bibnamefont {Pop}}, \ and\ \bibinfo {author}
  {\bibfnamefont {L.}~\bibnamefont {Shi}},\ }\bibfield  {title} {\emph
  {\bibinfo {title} {Nanoscale thermal transport},}\ }\href@noop {} {\bibfield
  {journal} {\bibinfo  {journal} {Appl. Phys. Rev.}\ }\textbf {\bibinfo
  {volume} {1}},\ \bibinfo {pages} {011305} (\bibinfo {year}
  {2014})}\BibitemShut {NoStop}%
\bibitem [{\citenamefont {Minnich}(2015)}]{Minnich2015}%
  \BibitemOpen
  \bibfield  {author} {\bibinfo {author} {\bibfnamefont {A.~J.}\ \bibnamefont
  {Minnich}},\ }\bibfield  {title} {\emph {\bibinfo {title} {Advances in the
  measurement and computation of thermal phonon transport properties.}}\
  }\href@noop {} {\bibfield  {journal} {\bibinfo  {journal} {Nano Res.}\
  }\textbf {\bibinfo {volume} {27}},\ \bibinfo {pages} {053202} (\bibinfo
  {year} {2015})}\BibitemShut {NoStop}%
\bibitem [{\citenamefont {Cui}\ \emph {et~al.}(2017)\citenamefont {Cui},
  \citenamefont {Jeong}, \citenamefont {Hur}, \citenamefont {Matt},
  \citenamefont {Kl{\"o}ckner}, \citenamefont {Pauly}, \citenamefont {Nielaba},
  \citenamefont {Cuevas}, \citenamefont {Meyhofer},\ and\ \citenamefont
  {Reddy}}]{Cui2017}%
  \BibitemOpen
  \bibfield  {author} {\bibinfo {author} {\bibfnamefont {L.}~\bibnamefont
  {Cui}}, \bibinfo {author} {\bibfnamefont {W.}~\bibnamefont {Jeong}}, \bibinfo
  {author} {\bibfnamefont {S.}~\bibnamefont {Hur}}, \bibinfo {author}
  {\bibfnamefont {M.}~\bibnamefont {Matt}}, \bibinfo {author} {\bibfnamefont
  {J.~C.}\ \bibnamefont {Kl{\"o}ckner}}, \bibinfo {author} {\bibfnamefont
  {F.}~\bibnamefont {Pauly}}, \bibinfo {author} {\bibfnamefont
  {P.}~\bibnamefont {Nielaba}}, \bibinfo {author} {\bibfnamefont {J.~C.}\
  \bibnamefont {Cuevas}}, \bibinfo {author} {\bibfnamefont {E.}~\bibnamefont
  {Meyhofer}}, \ and\ \bibinfo {author} {\bibfnamefont {P.}~\bibnamefont
  {Reddy}},\ }\bibfield  {title} {\emph {\bibinfo {title} {Quantized thermal
  transport in single-atom junctions},}\ }\href@noop {} {\bibfield  {journal}
  {\bibinfo  {journal} {Science}\ }\textbf {\bibinfo {volume} {355}},\ \bibinfo
  {pages} {1192} (\bibinfo {year} {2017})}\BibitemShut {NoStop}%
\bibitem [{\citenamefont {Mosso}\ \emph {et~al.}(2017)\citenamefont {Mosso},
  \citenamefont {Drechsler}, \citenamefont {Menges}, \citenamefont {Nirmalraj},
  \citenamefont {Karg}, \citenamefont {Riel},\ and\ \citenamefont
  {Gotsmann}}]{Mosso2017}%
  \BibitemOpen
  \bibfield  {author} {\bibinfo {author} {\bibfnamefont {N.}~\bibnamefont
  {Mosso}}, \bibinfo {author} {\bibfnamefont {U.}~\bibnamefont {Drechsler}},
  \bibinfo {author} {\bibfnamefont {F.}~\bibnamefont {Menges}}, \bibinfo
  {author} {\bibfnamefont {P.}~\bibnamefont {Nirmalraj}}, \bibinfo {author}
  {\bibfnamefont {S.}~\bibnamefont {Karg}}, \bibinfo {author} {\bibfnamefont
  {H.}~\bibnamefont {Riel}}, \ and\ \bibinfo {author} {\bibfnamefont
  {B.}~\bibnamefont {Gotsmann}},\ }\bibfield  {title} {\emph {\bibinfo {title}
  {Heat transport through atomic contacts},}\ }\href@noop {} {\bibfield
  {journal} {\bibinfo  {journal} {Nat. Nanotechnol.}\ }\textbf {\bibinfo
  {volume} {12}},\ \bibinfo {pages} {430} (\bibinfo {year} {2017})}\BibitemShut
  {NoStop}%
\bibitem [{\citenamefont {Cui}\ \emph {et~al.}(2019)\citenamefont {Cui},
  \citenamefont {Hur}, \citenamefont {Akbar}, \citenamefont {Kl\"ockner},
  \citenamefont {Jeong}, \citenamefont {Pauly},\ and\ \citenamefont
  {Jang}}]{Cui2019}%
  \BibitemOpen
  \bibfield  {author} {\bibinfo {author} {\bibfnamefont {L.}~\bibnamefont
  {Cui}}, \bibinfo {author} {\bibfnamefont {S.}~\bibnamefont {Hur}}, \bibinfo
  {author} {\bibfnamefont {Z.~A.}\ \bibnamefont {Akbar}}, \bibinfo {author}
  {\bibfnamefont {J.~C.}\ \bibnamefont {Kl\"ockner}}, \bibinfo {author}
  {\bibfnamefont {W.}~\bibnamefont {Jeong}}, \bibinfo {author} {\bibfnamefont
  {F.}~\bibnamefont {Pauly}}, \ and\ \bibinfo {author} {\bibfnamefont {S.-Y.}\
  \bibnamefont {Jang}},\ }\bibfield  {title} {\emph {\bibinfo {title} {Thermal
  conductance of single-molecule junctions},}\ }\href@noop {} {\bibfield
  {journal} {\bibinfo  {journal} {Nature}\ }\textbf {\bibinfo {volume} {572}},\
  \bibinfo {pages} {628} (\bibinfo {year} {2019})}\BibitemShut {NoStop}%
\bibitem [{\citenamefont {Mosso}\ \emph {et~al.}(2019)\citenamefont {Mosso},
  \citenamefont {Sadeghi}, \citenamefont {Gemma}, \citenamefont {Sangtarash},
  \citenamefont {Drechsler}, \citenamefont {Lambert},\ and\ \citenamefont
  {Gotsmann}}]{Mosso2019}%
  \BibitemOpen
  \bibfield  {author} {\bibinfo {author} {\bibfnamefont {N.}~\bibnamefont
  {Mosso}}, \bibinfo {author} {\bibfnamefont {H.}~\bibnamefont {Sadeghi}},
  \bibinfo {author} {\bibfnamefont {A.}~\bibnamefont {Gemma}}, \bibinfo
  {author} {\bibfnamefont {S.}~\bibnamefont {Sangtarash}}, \bibinfo {author}
  {\bibfnamefont {U.}~\bibnamefont {Drechsler}}, \bibinfo {author}
  {\bibfnamefont {C.}~\bibnamefont {Lambert}}, \ and\ \bibinfo {author}
  {\bibfnamefont {B.}~\bibnamefont {Gotsmann}},\ }\bibfield  {title} {\emph
  {\bibinfo {title} {Thermal Transport through Single-Molecule Junctions},}\
  }\href@noop {} {\bibfield  {journal} {\bibinfo  {journal} {Nano Lett.}\
  }\textbf {\bibinfo {volume} {19}},\ \bibinfo {pages} {7614} (\bibinfo {year}
  {2019})}\BibitemShut {NoStop}%
\bibitem [{\citenamefont {Agra{\"i}t}\ \emph {et~al.}(2003)\citenamefont
  {Agra{\"i}t}, \citenamefont {Yeyati},\ and\ \citenamefont {{van
  Ruitenbeek}}}]{Agrait2003}%
  \BibitemOpen
  \bibfield  {author} {\bibinfo {author} {\bibfnamefont {N.}~\bibnamefont
  {Agra{\"i}t}}, \bibinfo {author} {\bibfnamefont {A.~L.}\ \bibnamefont
  {Yeyati}}, \ and\ \bibinfo {author} {\bibfnamefont {J.~M.}\ \bibnamefont
  {{van Ruitenbeek}}},\ }\bibfield  {title} {\emph {\bibinfo {title} {Quantum
  properties of atomic-sized conductors},}\ }\href@noop {} {\bibfield
  {journal} {\bibinfo  {journal} {Phys. Rep.}\ }\textbf {\bibinfo {volume}
  {377}},\ \bibinfo {pages} {81} (\bibinfo {year} {2003})}\BibitemShut
  {NoStop}%
\bibitem [{\citenamefont {Cuevas}\ and\ \citenamefont
  {Scheer}(2017)}]{Cuevas2017}%
  \BibitemOpen
  \bibfield  {author} {\bibinfo {author} {\bibfnamefont {J.~C.}\ \bibnamefont
  {Cuevas}}\ and\ \bibinfo {author} {\bibfnamefont {E.}~\bibnamefont
  {Scheer}},\ }\href@noop {} {\emph {\bibinfo {title} {Molecular Electronics:
  An Introduction to Theory and Experiment}}},\ \bibinfo {edition} {2nd}\ ed.\
  (\bibinfo  {publisher} {World Scientific},\ \bibinfo {address} {Singapore},\
  \bibinfo {year} {2017})\BibitemShut {NoStop}%
\bibitem [{\citenamefont {Krans}\ \emph {et~al.}(1995)\citenamefont {Krans},
  \citenamefont {van Ruitenbeek}, \citenamefont {Fisun}, \citenamefont
  {Yanson},\ and\ \citenamefont {de~Jongh}}]{Krans1995}%
  \BibitemOpen
  \bibfield  {author} {\bibinfo {author} {\bibfnamefont {J.~M.}\ \bibnamefont
  {Krans}}, \bibinfo {author} {\bibfnamefont {J.~M.}\ \bibnamefont {van
  Ruitenbeek}}, \bibinfo {author} {\bibfnamefont {V.~V.}\ \bibnamefont
  {Fisun}}, \bibinfo {author} {\bibfnamefont {I.~K.}\ \bibnamefont {Yanson}}, \
  and\ \bibinfo {author} {\bibfnamefont {L.~J.}\ \bibnamefont {de~Jongh}},\
  }\bibfield  {title} {\emph {\bibinfo {title} {The signature of conductance
  quantization in metallic point contacts},}\ }\href@noop {} {\bibfield
  {journal} {\bibinfo  {journal} {Nature}\ }\textbf {\bibinfo {volume} {375}},\
  \bibinfo {pages} {767} (\bibinfo {year} {1995})}\BibitemShut {NoStop}%
\bibitem [{\citenamefont {Scheer}\ \emph {et~al.}(1998)\citenamefont {Scheer},
  \citenamefont {Agra{\"i}t}, \citenamefont {Cuevas}, \citenamefont {Yeyati},
  \citenamefont {Ludoph}, \citenamefont {Mart{\'i}n-Rodero}, \citenamefont
  {Bollinger}, \citenamefont {van Ruitenbeek},\ and\ \citenamefont
  {Urbina}}]{Scheer1998}%
  \BibitemOpen
  \bibfield  {author} {\bibinfo {author} {\bibfnamefont {E.}~\bibnamefont
  {Scheer}}, \bibinfo {author} {\bibfnamefont {N.}~\bibnamefont {Agra{\"i}t}},
  \bibinfo {author} {\bibfnamefont {J.~C.}\ \bibnamefont {Cuevas}}, \bibinfo
  {author} {\bibfnamefont {A.~L.}\ \bibnamefont {Yeyati}}, \bibinfo {author}
  {\bibfnamefont {B.}~\bibnamefont {Ludoph}}, \bibinfo {author} {\bibfnamefont
  {A.}~\bibnamefont {Mart{\'i}n-Rodero}}, \bibinfo {author} {\bibfnamefont
  {G.~R.}\ \bibnamefont {Bollinger}}, \bibinfo {author} {\bibfnamefont {J.~M.}\
  \bibnamefont {van Ruitenbeek}}, \ and\ \bibinfo {author} {\bibfnamefont
  {C.}~\bibnamefont {Urbina}},\ }\bibfield  {title} {\emph {\bibinfo {title}
  {The signature of chemical valence in the electrical conduction through a
  single-atom contact},}\ }\href@noop {} {\bibfield  {journal} {\bibinfo
  {journal} {Nature}\ }\textbf {\bibinfo {volume} {394}},\ \bibinfo {pages}
  {154} (\bibinfo {year} {1998})}\BibitemShut {NoStop}%
\bibitem [{\citenamefont {van~den Brom}\ and\ \citenamefont {van
  Ruitenbeek}(1999)}]{Brom1999}%
  \BibitemOpen
  \bibfield  {author} {\bibinfo {author} {\bibfnamefont {H.~E.}\ \bibnamefont
  {van~den Brom}}\ and\ \bibinfo {author} {\bibfnamefont {J.~M.}\ \bibnamefont
  {van Ruitenbeek}},\ }\bibfield  {title} {\emph {\bibinfo {title} {Quantum
  suppression of shot noise in atom-size metallic contacts},}\ }\href@noop {}
  {\bibfield  {journal} {\bibinfo  {journal} {Phys. Rev. Lett.}\ }\textbf
  {\bibinfo {volume} {82}},\ \bibinfo {pages} {1526} (\bibinfo {year}
  {1999})}\BibitemShut {NoStop}%
\bibitem [{\citenamefont {Wheeler}\ \emph {et~al.}(2010)\citenamefont
  {Wheeler}, \citenamefont {Russom}, \citenamefont {Evans}, \citenamefont
  {King},\ and\ \citenamefont {Natelson}}]{Wheeler2010}%
  \BibitemOpen
  \bibfield  {author} {\bibinfo {author} {\bibfnamefont {P.~J.}\ \bibnamefont
  {Wheeler}}, \bibinfo {author} {\bibfnamefont {J.~N.}\ \bibnamefont {Russom}},
  \bibinfo {author} {\bibfnamefont {K.}~\bibnamefont {Evans}}, \bibinfo
  {author} {\bibfnamefont {N.~S.}\ \bibnamefont {King}}, \ and\ \bibinfo
  {author} {\bibfnamefont {D.}~\bibnamefont {Natelson}},\ }\bibfield  {title}
  {\emph {\bibinfo {title} {Shot noise suppression at room temperature in
  atomic-scale {Au} junctions},}\ }\href@noop {} {\bibfield  {journal}
  {\bibinfo  {journal} {Nano Lett.}\ }\textbf {\bibinfo {volume} {10}},\
  \bibinfo {pages} {1287} (\bibinfo {year} {2010})}\BibitemShut {NoStop}%
\bibitem [{\citenamefont {Chen}\ \emph {et~al.}(2012)\citenamefont {Chen},
  \citenamefont {Wheeler},\ and\ \citenamefont {Natelson}}]{Chen2012}%
  \BibitemOpen
  \bibfield  {author} {\bibinfo {author} {\bibfnamefont {R.}~\bibnamefont
  {Chen}}, \bibinfo {author} {\bibfnamefont {P.~J.}\ \bibnamefont {Wheeler}}, \
  and\ \bibinfo {author} {\bibfnamefont {D.}~\bibnamefont {Natelson}},\
  }\bibfield  {title} {\emph {\bibinfo {title} {Excess noise in {STM}-style
  break junctions at room temperature},}\ }\href@noop {} {\bibfield  {journal}
  {\bibinfo  {journal} {Phys. Rev. B}\ }\textbf {\bibinfo {volume} {85}},\
  \bibinfo {pages} {235455} (\bibinfo {year} {2012})}\BibitemShut {NoStop}%
\bibitem [{\citenamefont {Vardimon}\ \emph {et~al.}(2016)\citenamefont
  {Vardimon}, \citenamefont {Matt}, \citenamefont {Nielaba}, \citenamefont
  {Cuevas},\ and\ \citenamefont {Tal}}]{Vardimon2016}%
  \BibitemOpen
  \bibfield  {author} {\bibinfo {author} {\bibfnamefont {R.}~\bibnamefont
  {Vardimon}}, \bibinfo {author} {\bibfnamefont {M.}~\bibnamefont {Matt}},
  \bibinfo {author} {\bibfnamefont {P.}~\bibnamefont {Nielaba}}, \bibinfo
  {author} {\bibfnamefont {J.~C.}\ \bibnamefont {Cuevas}}, \ and\ \bibinfo
  {author} {\bibfnamefont {O.}~\bibnamefont {Tal}},\ }\bibfield  {title} {\emph
  {\bibinfo {title} {Orbital origin of the electrical conduction in
  ferromagnetic atomic-size contacts: Insights from shot noise measurements and
  theoretical simulations},}\ }\href@noop {} {\bibfield  {journal} {\bibinfo
  {journal} {Phys. Rev. B}\ }\textbf {\bibinfo {volume} {93}},\ \bibinfo
  {pages} {085439} (\bibinfo {year} {2016})}\BibitemShut {NoStop}%
\bibitem [{\citenamefont {Guhr}\ \emph {et~al.}(2007)\citenamefont {Guhr},
  \citenamefont {Rettinger}, \citenamefont {Boneberg}, \citenamefont {Erbe},
  \citenamefont {Leiderer},\ and\ \citenamefont {Scheer}}]{Guhr2007}%
  \BibitemOpen
  \bibfield  {author} {\bibinfo {author} {\bibfnamefont {D.~C.}\ \bibnamefont
  {Guhr}}, \bibinfo {author} {\bibfnamefont {D.}~\bibnamefont {Rettinger}},
  \bibinfo {author} {\bibfnamefont {J.}~\bibnamefont {Boneberg}}, \bibinfo
  {author} {\bibfnamefont {A.}~\bibnamefont {Erbe}}, \bibinfo {author}
  {\bibfnamefont {P.}~\bibnamefont {Leiderer}}, \ and\ \bibinfo {author}
  {\bibfnamefont {E.}~\bibnamefont {Scheer}},\ }\bibfield  {title} {\emph
  {\bibinfo {title} {Influence of laser light on electronic transport through
  atomic-size contacts},}\ }\href@noop {} {\bibfield  {journal} {\bibinfo
  {journal} {Phys. Rev. Lett.}\ }\textbf {\bibinfo {volume} {99}},\ \bibinfo
  {pages} {086801} (\bibinfo {year} {2007})}\BibitemShut {NoStop}%
\bibitem [{\citenamefont {Viljas}\ and\ \citenamefont
  {Cuevas}(2007)}]{Viljas2007}%
  \BibitemOpen
  \bibfield  {author} {\bibinfo {author} {\bibfnamefont {J.~K.}\ \bibnamefont
  {Viljas}}\ and\ \bibinfo {author} {\bibfnamefont {J.~C.}\ \bibnamefont
  {Cuevas}},\ }\bibfield  {title} {\emph {\bibinfo {title} {Role of electronic
  structure in photoassisted transport through atomic-sized contacts},}\
  }\href@noop {} {\bibfield  {journal} {\bibinfo  {journal} {Phys. Rev. B}\
  }\textbf {\bibinfo {volume} {75}},\ \bibinfo {pages} {075406} (\bibinfo
  {year} {2007})}\BibitemShut {NoStop}%
\bibitem [{\citenamefont {Ittah}\ \emph {et~al.}(2009)\citenamefont {Ittah},
  \citenamefont {Noy}, \citenamefont {Yutsis},\ and\ \citenamefont
  {Selzer}}]{Ittah2009}%
  \BibitemOpen
  \bibfield  {author} {\bibinfo {author} {\bibfnamefont {N.}~\bibnamefont
  {Ittah}}, \bibinfo {author} {\bibfnamefont {G.}~\bibnamefont {Noy}}, \bibinfo
  {author} {\bibfnamefont {I.}~\bibnamefont {Yutsis}}, \ and\ \bibinfo {author}
  {\bibfnamefont {Y.}~\bibnamefont {Selzer}},\ }\bibfield  {title} {\emph
  {\bibinfo {title} {Measurement of electronic transport through $1G_0$ gold
  contacts under laser irradiation},}\ }\href@noop {} {\bibfield  {journal}
  {\bibinfo  {journal} {Nano Lett.}\ }\textbf {\bibinfo {volume} {9}},\
  \bibinfo {pages} {1615} (\bibinfo {year} {2009})}\BibitemShut {NoStop}%
\bibitem [{\citenamefont {Ward}\ \emph {et~al.}(2010)\citenamefont {Ward},
  \citenamefont {H{\"u}ser}, \citenamefont {Pauly}, \citenamefont {Cuevas},\
  and\ \citenamefont {Natelson}}]{Ward2010}%
  \BibitemOpen
  \bibfield  {author} {\bibinfo {author} {\bibfnamefont {D.~R.}\ \bibnamefont
  {Ward}}, \bibinfo {author} {\bibfnamefont {F.}~\bibnamefont {H{\"u}ser}},
  \bibinfo {author} {\bibfnamefont {F.}~\bibnamefont {Pauly}}, \bibinfo
  {author} {\bibfnamefont {J.~C.}\ \bibnamefont {Cuevas}}, \ and\ \bibinfo
  {author} {\bibfnamefont {D.}~\bibnamefont {Natelson}},\ }\bibfield  {title}
  {\emph {\bibinfo {title} {Optical rectification and field enhancement in a
  plasmonic nanogap},}\ }\href@noop {} {\bibfield  {journal} {\bibinfo
  {journal} {Nat. Nanotechnol.}\ }\textbf {\bibinfo {volume} {5}},\ \bibinfo
  {pages} {732} (\bibinfo {year} {2010})}\BibitemShut {NoStop}%
\bibitem [{\citenamefont {Ludoph}\ and\ \citenamefont {van
  Ruitenbeek}(1999)}]{Ludoph1999}%
  \BibitemOpen
  \bibfield  {author} {\bibinfo {author} {\bibfnamefont {B.}~\bibnamefont
  {Ludoph}}\ and\ \bibinfo {author} {\bibfnamefont {J.~M.}\ \bibnamefont {van
  Ruitenbeek}},\ }\bibfield  {title} {\emph {\bibinfo {title} {Thermopower of
  atomic-size metallic contacts},}\ }\href@noop {} {\bibfield  {journal}
  {\bibinfo  {journal} {Phys. Rev. B}\ }\textbf {\bibinfo {volume} {59}},\
  \bibinfo {pages} {12290} (\bibinfo {year} {1999})}\BibitemShut {NoStop}%
\bibitem [{\citenamefont {Tsutsui}\ \emph {et~al.}(2013)\citenamefont
  {Tsutsui}, \citenamefont {Morikawa}, \citenamefont {Arima},\ and\
  \citenamefont {Taniguchi}}]{Tsutsui2013}%
  \BibitemOpen
  \bibfield  {author} {\bibinfo {author} {\bibfnamefont {M.}~\bibnamefont
  {Tsutsui}}, \bibinfo {author} {\bibfnamefont {T.}~\bibnamefont {Morikawa}},
  \bibinfo {author} {\bibfnamefont {A.}~\bibnamefont {Arima}}, \ and\ \bibinfo
  {author} {\bibfnamefont {M.}~\bibnamefont {Taniguchi}},\ }\bibfield  {title}
  {\emph {\bibinfo {title} {Thermoelectricity in atom-sized junctions at room
  temperatures},}\ }\href@noop {} {\bibfield  {journal} {\bibinfo  {journal}
  {Sci. Rep.}\ }\textbf {\bibinfo {volume} {3}},\ \bibinfo {pages} {3326}
  (\bibinfo {year} {2013})}\BibitemShut {NoStop}%
\bibitem [{\citenamefont {Evangeli}\ \emph {et~al.}(2015)\citenamefont
  {Evangeli}, \citenamefont {Matt}, \citenamefont {Rinc{\'o}n-Garc{\'i}a},
  \citenamefont {Pauly}, \citenamefont {Nielaba}, \citenamefont
  {Rubio-Bollinger}, \citenamefont {Cuevas},\ and\ \citenamefont
  {Agra{\"i}t}}]{Evangeli2015}%
  \BibitemOpen
  \bibfield  {author} {\bibinfo {author} {\bibfnamefont {C.}~\bibnamefont
  {Evangeli}}, \bibinfo {author} {\bibfnamefont {M.}~\bibnamefont {Matt}},
  \bibinfo {author} {\bibfnamefont {L.}~\bibnamefont {Rinc{\'o}n-Garc{\'i}a}},
  \bibinfo {author} {\bibfnamefont {F.}~\bibnamefont {Pauly}}, \bibinfo
  {author} {\bibfnamefont {P.}~\bibnamefont {Nielaba}}, \bibinfo {author}
  {\bibfnamefont {G.}~\bibnamefont {Rubio-Bollinger}}, \bibinfo {author}
  {\bibfnamefont {J.~C.}\ \bibnamefont {Cuevas}}, \ and\ \bibinfo {author}
  {\bibfnamefont {N.}~\bibnamefont {Agra{\"i}t}},\ }\bibfield  {title} {\emph
  {\bibinfo {title} {Quantum thermopower of metallic atomic-size contacts at
  room temperature},}\ }\href@noop {} {\bibfield  {journal} {\bibinfo
  {journal} {Nano Lett.}\ }\textbf {\bibinfo {volume} {15}},\ \bibinfo {pages}
  {1006} (\bibinfo {year} {2015})}\BibitemShut {NoStop}%
\bibitem [{\citenamefont {Ofarim}\ \emph {et~al.}(2016)\citenamefont {Ofarim},
  \citenamefont {Kopp}, \citenamefont {M\"oller}, \citenamefont {Martin},
  \citenamefont {Boneberg}, \citenamefont {Leiderer},\ and\ \citenamefont
  {Scheer}}]{Ofarim2016}%
  \BibitemOpen
  \bibfield  {author} {\bibinfo {author} {\bibfnamefont {A.}~\bibnamefont
  {Ofarim}}, \bibinfo {author} {\bibfnamefont {B.}~\bibnamefont {Kopp}},
  \bibinfo {author} {\bibfnamefont {T.}~\bibnamefont {M\"oller}}, \bibinfo
  {author} {\bibfnamefont {L.}~\bibnamefont {Martin}}, \bibinfo {author}
  {\bibfnamefont {J.}~\bibnamefont {Boneberg}}, \bibinfo {author}
  {\bibfnamefont {P.}~\bibnamefont {Leiderer}}, \ and\ \bibinfo {author}
  {\bibfnamefont {E.}~\bibnamefont {Scheer}},\ }\bibfield  {title} {\emph
  {\bibinfo {title} {Thermo-voltage measurements of atomic contacts at low
  temperature},}\ }\href@noop {} {\bibfield  {journal} {\bibinfo  {journal}
  {Beilstein J. Nanotechnol.}\ }\textbf {\bibinfo {volume} {7}},\ \bibinfo
  {pages} {767} (\bibinfo {year} {2016})}\BibitemShut {NoStop}%
\bibitem [{\citenamefont {Lee}\ \emph {et~al.}(2013)\citenamefont {Lee},
  \citenamefont {Kim}, \citenamefont {Jeong}, \citenamefont {Zotti},
  \citenamefont {Pauly}, \citenamefont {Cuevas},\ and\ \citenamefont
  {Reddy}}]{Lee2013}%
  \BibitemOpen
  \bibfield  {author} {\bibinfo {author} {\bibfnamefont {W.}~\bibnamefont
  {Lee}}, \bibinfo {author} {\bibfnamefont {K.}~\bibnamefont {Kim}}, \bibinfo
  {author} {\bibfnamefont {W.}~\bibnamefont {Jeong}}, \bibinfo {author}
  {\bibfnamefont {L.~A.}\ \bibnamefont {Zotti}}, \bibinfo {author}
  {\bibfnamefont {F.}~\bibnamefont {Pauly}}, \bibinfo {author} {\bibfnamefont
  {J.~C.}\ \bibnamefont {Cuevas}}, \ and\ \bibinfo {author} {\bibfnamefont
  {P.}~\bibnamefont {Reddy}},\ }\bibfield  {title} {\emph {\bibinfo {title}
  {Heat dissipation in atomic-scale junctions},}\ }\href@noop {} {\bibfield
  {journal} {\bibinfo  {journal} {Nature}\ }\textbf {\bibinfo {volume} {498}},\
  \bibinfo {pages} {209} (\bibinfo {year} {2013})}\BibitemShut {NoStop}%
\bibitem [{\citenamefont {Zotti}\ \emph {et~al.}(2014)\citenamefont {Zotti},
  \citenamefont {B\"urkle}, \citenamefont {Pauly}, \citenamefont {Lee},
  \citenamefont {Kim}, \citenamefont {Jeong}, \citenamefont {Asai},
  \citenamefont {Reddy},\ and\ \citenamefont {Cuevas}}]{Zotti2014}%
  \BibitemOpen
  \bibfield  {author} {\bibinfo {author} {\bibfnamefont {L.~A.}\ \bibnamefont
  {Zotti}}, \bibinfo {author} {\bibfnamefont {M.}~\bibnamefont {B\"urkle}},
  \bibinfo {author} {\bibfnamefont {F.}~\bibnamefont {Pauly}}, \bibinfo
  {author} {\bibfnamefont {W.}~\bibnamefont {Lee}}, \bibinfo {author}
  {\bibfnamefont {K.}~\bibnamefont {Kim}}, \bibinfo {author} {\bibfnamefont
  {W.}~\bibnamefont {Jeong}}, \bibinfo {author} {\bibfnamefont
  {Y.}~\bibnamefont {Asai}}, \bibinfo {author} {\bibfnamefont {P.}~\bibnamefont
  {Reddy}}, \ and\ \bibinfo {author} {\bibfnamefont {J.~C.}\ \bibnamefont
  {Cuevas}},\ }\bibfield  {title} {\emph {\bibinfo {title} {Heat dissipation
  and its relation to thermopower in single-molecule junctions},}\ }\href@noop
  {} {\bibfield  {journal} {\bibinfo  {journal} {New J. Phys.}\ }\textbf
  {\bibinfo {volume} {16}},\ \bibinfo {pages} {015004} (\bibinfo {year}
  {2014})}\BibitemShut {NoStop}%
\bibitem [{\citenamefont {Cui}\ \emph {et~al.}(2018)\citenamefont {Cui},
  \citenamefont {Miao}, \citenamefont {Wang}, \citenamefont {Thompson},
  \citenamefont {Zotti}, \citenamefont {Cuevas}, \citenamefont {Meyhofer},\
  and\ \citenamefont {Reddy}}]{Cui2018}%
  \BibitemOpen
  \bibfield  {author} {\bibinfo {author} {\bibfnamefont {L.}~\bibnamefont
  {Cui}}, \bibinfo {author} {\bibfnamefont {R.}~\bibnamefont {Miao}}, \bibinfo
  {author} {\bibfnamefont {K.}~\bibnamefont {Wang}}, \bibinfo {author}
  {\bibfnamefont {D.}~\bibnamefont {Thompson}}, \bibinfo {author}
  {\bibfnamefont {L.~A.}\ \bibnamefont {Zotti}}, \bibinfo {author}
  {\bibfnamefont {J.~C.}\ \bibnamefont {Cuevas}}, \bibinfo {author}
  {\bibfnamefont {E.}~\bibnamefont {Meyhofer}}, \ and\ \bibinfo {author}
  {\bibfnamefont {P.}~\bibnamefont {Reddy}},\ }\bibfield  {title} {\emph
  {\bibinfo {title} {Peltier cooling in molecular junctions},}\ }\href@noop {}
  {\bibfield  {journal} {\bibinfo  {journal} {Nat. Nanotechnol.}\ }\textbf
  {\bibinfo {volume} {13}},\ \bibinfo {pages} {122} (\bibinfo {year}
  {2018})}\BibitemShut {NoStop}%
\bibitem [{\citenamefont {Kl\"ockner}\ \emph {et~al.}(2017)\citenamefont
  {Kl\"ockner}, \citenamefont {Matt}, \citenamefont {Nielaba}, \citenamefont
  {Pauly},\ and\ \citenamefont {Cuevas}}]{Klockner2017}%
  \BibitemOpen
  \bibfield  {author} {\bibinfo {author} {\bibfnamefont {J.~C.}\ \bibnamefont
  {Kl\"ockner}}, \bibinfo {author} {\bibfnamefont {M.}~\bibnamefont {Matt}},
  \bibinfo {author} {\bibfnamefont {P.}~\bibnamefont {Nielaba}}, \bibinfo
  {author} {\bibfnamefont {F.}~\bibnamefont {Pauly}}, \ and\ \bibinfo {author}
  {\bibfnamefont {J.~C.}\ \bibnamefont {Cuevas}},\ }\bibfield  {title} {\emph
  {\bibinfo {title} {Thermal conductance of metallic atomic-size contacts:
  Phonon transport and Wiedemann-Franz law},}\ }\href@noop {} {\bibfield
  {journal} {\bibinfo  {journal} {Phys. Rev. B}\ }\textbf {\bibinfo {volume}
  {96}},\ \bibinfo {pages} {205405} (\bibinfo {year} {2017})}\BibitemShut
  {NoStop}%
\bibitem [{\citenamefont {B\"urkle}\ and\ \citenamefont
  {Asai}(2018)}]{Burkle2018}%
  \BibitemOpen
  \bibfield  {author} {\bibinfo {author} {\bibfnamefont {M.}~\bibnamefont
  {B\"urkle}}\ and\ \bibinfo {author} {\bibfnamefont {Y.}~\bibnamefont
  {Asai}},\ }\bibfield  {title} {\emph {\bibinfo {title} {How to probe the
  limits of the Wiedemann-Franz law at nanoscale},}\ }\href@noop {}
  {\bibfield  {journal} {\bibinfo  {journal} {Nano Lett.}\ }\textbf {\bibinfo
  {volume} {18}},\ \bibinfo {pages} {7358} (\bibinfo {year}
  {2018})}\BibitemShut {NoStop}%
\bibitem [{\citenamefont {M\"ohrle}\ \emph {et~al.}(2019)\citenamefont
  {M\"ohrle}, \citenamefont {M\"uller}, \citenamefont {Matt}, \citenamefont
  {Nielaba},\ and\ \citenamefont {Pauly}}]{Moehrle2019}%
  \BibitemOpen
  \bibfield  {author} {\bibinfo {author} {\bibfnamefont {D.~O.}\ \bibnamefont
  {M\"ohrle}}, \bibinfo {author} {\bibfnamefont {F.}~\bibnamefont {M\"uller}},
  \bibinfo {author} {\bibfnamefont {M.}~\bibnamefont {Matt}}, \bibinfo {author}
  {\bibfnamefont {P.}~\bibnamefont {Nielaba}}, \ and\ \bibinfo {author}
  {\bibfnamefont {F.}~\bibnamefont {Pauly}},\ }\bibfield  {title} {\emph
  {\bibinfo {title} {Statistical analysis of electronic and phononic transport
  simulations of metallic atomic contacts},}\ }\href@noop {} {\bibfield
  {journal} {\bibinfo  {journal} {Phys. Rev. B}\ }\textbf {\bibinfo {volume}
  {100}},\ \bibinfo {pages} {125433} (\bibinfo {year} {2019})}\BibitemShut
  {NoStop}%
\bibitem [{\citenamefont {Jain}\ and\ \citenamefont
  {McGaughey}(2016)}]{Jain2016}%
  \BibitemOpen
  \bibfield  {author} {\bibinfo {author} {\bibfnamefont {A.}~\bibnamefont
  {Jain}}\ and\ \bibinfo {author} {\bibfnamefont {A.~J.~H.}\ \bibnamefont
  {McGaughey}},\ }\bibfield  {title} {\emph {\bibinfo {title} {Thermal
  transport by phonons and electrons in aluminum, silver, and gold from first
  principles},}\ }\href@noop {} {\bibfield  {journal} {\bibinfo  {journal}
  {Phys. Rev. B}\ }\textbf {\bibinfo {volume} {93}},\ \bibinfo {pages} {081206}
  (\bibinfo {year} {2016})}\BibitemShut {NoStop}%
\bibitem [{\citenamefont {Mingo}\ and\ \citenamefont {Yang}(2003)}]{Mingo2003}%
  \BibitemOpen
  \bibfield  {author} {\bibinfo {author} {\bibfnamefont {N.}~\bibnamefont
  {Mingo}}\ and\ \bibinfo {author} {\bibfnamefont {L.}~\bibnamefont {Yang}},\
  }\bibfield  {title} {\emph {\bibinfo {title} {Phonon transport in nanowires
  coated with an amorphous material: An atomistic Green's function approach},}\
  }\href@noop {} {\bibfield  {journal} {\bibinfo  {journal} {Phys. Rev. B}\
  }\textbf {\bibinfo {volume} {68}},\ \bibinfo {pages} {245406} (\bibinfo
  {year} {2003})}\BibitemShut {NoStop}%
\bibitem [{\citenamefont {Kl\"ockner}\ \emph {et~al.}(2018)\citenamefont
  {Kl\"ockner}, \citenamefont {Cuevas},\ and\ \citenamefont
  {Pauly}}]{Kloeckner2018}%
  \BibitemOpen
  \bibfield  {author} {\bibinfo {author} {\bibfnamefont {J.~C.}\ \bibnamefont
  {Kl\"ockner}}, \bibinfo {author} {\bibfnamefont {J.~C.}\ \bibnamefont
  {Cuevas}}, \ and\ \bibinfo {author} {\bibfnamefont {F.}~\bibnamefont
  {Pauly}},\ }\bibfield  {title} {\emph {\bibinfo {title} {Transmission
  eigenchannels for coherent phonon transport},}\ }\href@noop {} {\bibfield
  {journal} {\bibinfo  {journal} {Phys. Rev. B}\ }\textbf {\bibinfo {volume}
  {97}},\ \bibinfo {pages} {155432} (\bibinfo {year} {2018})}\BibitemShut
  {NoStop}%
\bibitem [{\citenamefont {Ashcroft}\ and\ \citenamefont
  {Mermin}(1976)}]{Ashcroft1976}%
  \BibitemOpen
  \bibfield  {author} {\bibinfo {author} {\bibfnamefont {N.~W.}\ \bibnamefont
  {Ashcroft}}\ and\ \bibinfo {author} {\bibfnamefont {N.~D.}\ \bibnamefont
  {Mermin}},\ }\href@noop {} {\emph {\bibinfo {title} {Solid State Physics}}}\
  (\bibinfo  {publisher} {Harcourt, Inc.},\ \bibinfo {address} {Orlando},\
  \bibinfo {year} {1976})\BibitemShut {NoStop}%
\bibitem [{\citenamefont {Campa\~n\'a}\ and\ \citenamefont
  {M\"user}(2006)}]{Campana2006}%
  \BibitemOpen
  \bibfield  {author} {\bibinfo {author} {\bibfnamefont {C.}~\bibnamefont
  {Campa\~n\'a}}\ and\ \bibinfo {author} {\bibfnamefont {M.~H.}\ \bibnamefont
  {M\"user}},\ }\bibfield  {title} {\emph {\bibinfo {title} {Practical Green's
  function approach to the simulation of elastic semi-infinite solids},}\
  }\href@noop {} {\bibfield  {journal} {\bibinfo  {journal} {Phys. Rev. B}\
  }\textbf {\bibinfo {volume} {74}},\ \bibinfo {pages} {075420} (\bibinfo
  {year} {2006})}\BibitemShut {NoStop}%
\bibitem [{\citenamefont {Kong}\ \emph {et~al.}(2009)\citenamefont {Kong},
  \citenamefont {Bartels}, \citenamefont {Campa{\~n}{\'a}}, \citenamefont
  {Denniston},\ and\ \citenamefont {M{\"u}ser}}]{Kong2009}%
  \BibitemOpen
  \bibfield  {author} {\bibinfo {author} {\bibfnamefont {L.~T.}\ \bibnamefont
  {Kong}}, \bibinfo {author} {\bibfnamefont {G.}~\bibnamefont {Bartels}},
  \bibinfo {author} {\bibfnamefont {C.}~\bibnamefont {Campa{\~n}{\'a}}},
  \bibinfo {author} {\bibfnamefont {C.}~\bibnamefont {Denniston}}, \ and\
  \bibinfo {author} {\bibfnamefont {M.~H.}\ \bibnamefont {M{\"u}ser}},\
  }\bibfield  {title} {\emph {\bibinfo {title} {Implementation of Green's
  function molecular dynamics: An extension to \textsc{Lammps}},}\ }\href@noop
  {} {\bibfield  {journal} {\bibinfo  {journal} {Comput. Phys. Commun.}\
  }\textbf {\bibinfo {volume} {180}},\ \bibinfo {pages} {1004} (\bibinfo {year}
  {2009})}\BibitemShut {NoStop}%
\bibitem [{\citenamefont {Kong}(2011)}]{Kong2011}%
  \BibitemOpen
  \bibfield  {author} {\bibinfo {author} {\bibfnamefont {L.~T.}\ \bibnamefont
  {Kong}},\ }\bibfield  {title} {\emph {\bibinfo {title} {Phonon dispersion
  measured directly from molecular dynamics simulations},}\ }\href@noop {}
  {\bibfield  {journal} {\bibinfo  {journal} {Comput. Phys. Commun.}\ }\textbf
  {\bibinfo {volume} {182}},\ \bibinfo {pages} {2201} (\bibinfo {year}
  {2011})}\BibitemShut {NoStop}%
\bibitem [{\citenamefont {Landau}\ and\ \citenamefont
  {Lifshitz}(1980)}]{Landau1980}%
  \BibitemOpen
  \bibfield  {author} {\bibinfo {author} {\bibfnamefont {L.~D.}\ \bibnamefont
  {Landau}}\ and\ \bibinfo {author} {\bibfnamefont {E.~M.}\ \bibnamefont
  {Lifshitz}},\ }\href@noop {} {\emph {\bibinfo {title} {Statistical
  Physics}}}\ (\bibinfo  {publisher} {Elsevier Science},\ \bibinfo {year}
  {1980})\BibitemShut {NoStop}%
\bibitem [{\citenamefont {Guinea}\ \emph {et~al.}(1983)\citenamefont {Guinea},
  \citenamefont {Tejedor}, \citenamefont {Flores},\ and\ \citenamefont
  {Louis}}]{Guinea1983}%
  \BibitemOpen
  \bibfield  {author} {\bibinfo {author} {\bibfnamefont {F.}~\bibnamefont
  {Guinea}}, \bibinfo {author} {\bibfnamefont {C.}~\bibnamefont {Tejedor}},
  \bibinfo {author} {\bibfnamefont {F.}~\bibnamefont {Flores}}, \ and\ \bibinfo
  {author} {\bibfnamefont {E.}~\bibnamefont {Louis}},\ }\bibfield  {title}
  {\emph {\bibinfo {title} {Effective two-dimensional {Hamiltonian} at
  surfaces},}\ }\href@noop {} {\bibfield  {journal} {\bibinfo  {journal} {Phys.
  Rev. B}\ }\textbf {\bibinfo {volume} {28}},\ \bibinfo {pages} {4397}
  (\bibinfo {year} {1983})}\BibitemShut {NoStop}%
\bibitem [{\citenamefont {Pauly}\ \emph {et~al.}(2008)\citenamefont {Pauly},
  \citenamefont {Viljas}, \citenamefont {Huniar}, \citenamefont {H\"afner},
  \citenamefont {Wohlthat}, \citenamefont {B\"urkle}, \citenamefont {Cuevas},\
  and\ \citenamefont {Sch\"on}}]{Pauly2008}%
  \BibitemOpen
  \bibfield  {author} {\bibinfo {author} {\bibfnamefont {F.}~\bibnamefont
  {Pauly}}, \bibinfo {author} {\bibfnamefont {J.~K.}\ \bibnamefont {Viljas}},
  \bibinfo {author} {\bibfnamefont {U.}~\bibnamefont {Huniar}}, \bibinfo
  {author} {\bibfnamefont {M.}~\bibnamefont {H\"afner}}, \bibinfo {author}
  {\bibfnamefont {S.}~\bibnamefont {Wohlthat}}, \bibinfo {author}
  {\bibfnamefont {M.}~\bibnamefont {B\"urkle}}, \bibinfo {author}
  {\bibfnamefont {J.~C.}\ \bibnamefont {Cuevas}}, \ and\ \bibinfo {author}
  {\bibfnamefont {G.}~\bibnamefont {Sch\"on}},\ }\bibfield  {title} {\emph
  {\bibinfo {title} {Cluster-based density-functional approach to quantum
  transport through molecular and atomic contacts},}\ }\href@noop {} {\bibfield
   {journal} {\bibinfo  {journal} {New J. Phys.}\ }\textbf {\bibinfo {volume}
  {10}},\ \bibinfo {pages} {125019} (\bibinfo {year} {2008})}\BibitemShut
  {NoStop}%
\bibitem [{\citenamefont {B\"urkle}\ \emph {et~al.}(2015)\citenamefont
  {B\"urkle}, \citenamefont {Hellmuth}, \citenamefont {Pauly},\ and\
  \citenamefont {Asai}}]{Buerkle2015}%
  \BibitemOpen
  \bibfield  {author} {\bibinfo {author} {\bibfnamefont {M.}~\bibnamefont
  {B\"urkle}}, \bibinfo {author} {\bibfnamefont {T.~J.}\ \bibnamefont
  {Hellmuth}}, \bibinfo {author} {\bibfnamefont {F.}~\bibnamefont {Pauly}}, \
  and\ \bibinfo {author} {\bibfnamefont {Y.}~\bibnamefont {Asai}},\ }\bibfield
  {title} {\emph {\bibinfo {title} {First-principles calculation of the
  thermoelectric figure of merit for [2,2]paracyclophane-based single-molecule
  junctions},}\ }\href@noop {} {\bibfield  {journal} {\bibinfo  {journal}
  {Phys. Rev. B}\ }\textbf {\bibinfo {volume} {91}},\ \bibinfo {pages} {165419}
  (\bibinfo {year} {2015})}\BibitemShut {NoStop}%
\bibitem [{\citenamefont {Plimpton}(1995)}]{Plimpton1995}%
  \BibitemOpen
  \bibfield  {author} {\bibinfo {author} {\bibfnamefont {S.}~\bibnamefont
  {Plimpton}},\ }\bibfield  {title} {\emph {\bibinfo {title} {Fast parallel
  algorithms for short-range molecular dynamics},}\ }\href@noop {} {\bibfield
  {journal} {\bibinfo  {journal} {J. Comput. Phys.}\ }\textbf {\bibinfo
  {volume} {117}},\ \bibinfo {pages} {1} (\bibinfo {year} {1995})}\BibitemShut
  {NoStop}%
\bibitem [{Lam()}]{Lammps}%
  \BibitemOpen
  \href@noop {} {}\bibinfo {note} {LAMMPS Molecular Dynamics Simulator,
  \url{http://lammps.sandia.gov}}\BibitemShut {NoStop}%
\bibitem [{\citenamefont {Finnis}\ and\ \citenamefont
  {Sinclair}(1984)}]{Finnis1984}%
  \BibitemOpen
  \bibfield  {author} {\bibinfo {author} {\bibfnamefont {M.~W.}\ \bibnamefont
  {Finnis}}\ and\ \bibinfo {author} {\bibfnamefont {J.~E.}\ \bibnamefont
  {Sinclair}},\ }\bibfield  {title} {\emph {\bibinfo {title} {A simple
  empirical $N$-body potential for transition metals},}\ }\href@noop {}
  {\bibfield  {journal} {\bibinfo  {journal} {Philos. Mag. A}\ }\textbf
  {\bibinfo {volume} {50}},\ \bibinfo {pages} {45} (\bibinfo {year}
  {1984})}\BibitemShut {NoStop}%
\bibitem [{\citenamefont {Ackland}\ \emph {et~al.}(1987)\citenamefont
  {Ackland}, \citenamefont {Tichy}, \citenamefont {Vitek},\ and\ \citenamefont
  {Finnis}}]{Ackland1987}%
  \BibitemOpen
  \bibfield  {author} {\bibinfo {author} {\bibfnamefont {G.~J.}\ \bibnamefont
  {Ackland}}, \bibinfo {author} {\bibfnamefont {G.}~\bibnamefont {Tichy}},
  \bibinfo {author} {\bibfnamefont {V.}~\bibnamefont {Vitek}}, \ and\ \bibinfo
  {author} {\bibfnamefont {M.~W.}\ \bibnamefont {Finnis}},\ }\bibfield  {title}
  {\emph {\bibinfo {title} {Simple $N$-body potentials for the noble metals and
  nickel},}\ }\href@noop {} {\bibfield  {journal} {\bibinfo  {journal} {Philos.
  Mag. A}\ }\textbf {\bibinfo {volume} {56}},\ \bibinfo {pages} {735} (\bibinfo
  {year} {1987})}\BibitemShut {NoStop}%
\bibitem [{\citenamefont {Lynn}\ \emph {et~al.}(1973)\citenamefont {Lynn},
  \citenamefont {Smith},\ and\ \citenamefont {Nicklow}}]{Lynn1973}%
  \BibitemOpen
  \bibfield  {author} {\bibinfo {author} {\bibfnamefont {J.~W.}\ \bibnamefont
  {Lynn}}, \bibinfo {author} {\bibfnamefont {H.~G.}\ \bibnamefont {Smith}}, \
  and\ \bibinfo {author} {\bibfnamefont {R.~M.}\ \bibnamefont {Nicklow}},\
  }\bibfield  {title} {\emph {\bibinfo {title} {Lattice dynamics of gold},}\
  }\href@noop {} {\bibfield  {journal} {\bibinfo  {journal} {Phys. Rev. B}\
  }\textbf {\bibinfo {volume} {8}},\ \bibinfo {pages} {3493} (\bibinfo {year}
  {1973})}\BibitemShut {NoStop}%
\bibitem [{\citenamefont {Mingo}(2006)}]{Mingo2006}%
  \BibitemOpen
  \bibfield  {author} {\bibinfo {author} {\bibfnamefont {N.}~\bibnamefont
  {Mingo}},\ }\bibfield  {title} {\emph {\bibinfo {title} {Anharmonic phonon
  flow through molecular-sized junctions},}\ }\href@noop {} {\bibfield
  {journal} {\bibinfo  {journal} {Phys. Rev. B}\ }\textbf {\bibinfo {volume}
  {74}},\ \bibinfo {pages} {125402} (\bibinfo {year} {2006})}\BibitemShut
  {NoStop}%
\bibitem [{\citenamefont {Smit}\ \emph {et~al.}(2001)\citenamefont {Smit},
  \citenamefont {Untiedt}, \citenamefont {Yanson},\ and\ \citenamefont {van
  Ruitenbeek}}]{Smit2001}%
  \BibitemOpen
  \bibfield  {author} {\bibinfo {author} {\bibfnamefont {R.~H.~M.}\
  \bibnamefont {Smit}}, \bibinfo {author} {\bibfnamefont {C.}~\bibnamefont
  {Untiedt}}, \bibinfo {author} {\bibfnamefont {A.~I.}\ \bibnamefont {Yanson}},
  \ and\ \bibinfo {author} {\bibfnamefont {J.~M.}\ \bibnamefont {van
  Ruitenbeek}},\ }\bibfield  {title} {\emph {\bibinfo {title} {Common Origin
  for Surface Reconstruction and the Formation of Chains of Metal Atoms},}\
  }\href@noop {} {\bibfield  {journal} {\bibinfo  {journal} {Phys. Rev. Lett.}\
  }\textbf {\bibinfo {volume} {87}},\ \bibinfo {pages} {266102} (\bibinfo
  {year} {2001})}\BibitemShut {NoStop}%
\bibitem [{\citenamefont {Yanson}\ \emph {et~al.}(1998)\citenamefont {Yanson},
  \citenamefont {Bollinger}, \citenamefont {van~den Brom}, \citenamefont
  {Agra{\"i}t},\ and\ \citenamefont {van Ruitenbeek}}]{Yanson1998}%
  \BibitemOpen
  \bibfield  {author} {\bibinfo {author} {\bibfnamefont {A.~I.}\ \bibnamefont
  {Yanson}}, \bibinfo {author} {\bibfnamefont {G.}\ \bibnamefont
  {Rubio-Bollinger}}, \bibinfo {author} {\bibfnamefont {H.~E.}\ \bibnamefont {van~den
  Brom}}, \bibinfo {author} {\bibfnamefont {N.}~\bibnamefont {Agra{\"i}t}}, \
  and\ \bibinfo {author} {\bibfnamefont {J.~M.}\ \bibnamefont {van
  Ruitenbeek}},\ }\bibfield  {title} {\emph {\bibinfo {title} {Formation and
  manipulation of a metallic wire of single gold atoms},}\ }\href@noop {}
  {\bibfield  {journal} {\bibinfo  {journal} {Nature}\ }\textbf {\bibinfo
  {volume} {395}},\ \bibinfo {pages} {783} (\bibinfo {year}
  {1998})}\BibitemShut {NoStop}%
\bibitem [{\citenamefont {Agra{\"i}t}\ \emph {et~al.}(2002)\citenamefont
  {Agra{\"i}t}, \citenamefont {Untiedt}, \citenamefont {Rubio-Bollinger},\ and\
  \citenamefont {Vieira}}]{Agrait2002}%
  \BibitemOpen
  \bibfield  {author} {\bibinfo {author} {\bibfnamefont {N.}~\bibnamefont
  {Agra{\"i}t}}, \bibinfo {author} {\bibfnamefont {C.}~\bibnamefont {Untiedt}},
  \bibinfo {author} {\bibfnamefont {G.}~\bibnamefont {Rubio-Bollinger}}, \ and\
  \bibinfo {author} {\bibfnamefont {S.}~\bibnamefont {Vieira}},\ }\bibfield
  {title} {\emph {\bibinfo {title} {Onset of Energy Dissipation in Ballistic
  Atomic Wires},}\ }\href@noop {} {\bibfield  {journal} {\bibinfo  {journal}
  {Phys. Rev. Lett.}\ }\textbf {\bibinfo {volume} {88}},\ \bibinfo {pages}
  {216803} (\bibinfo {year} {2002})}\BibitemShut {NoStop}%
\end{thebibliography}
\end{document}